\documentclass{elsart}
\usepackage{ifpdf}
\usepackage{graphicx,amssymb,lineno,epsfig,subfig}
\usepackage{subfig}
\usepackage{indentfirst}

\ifpdf
\usepackage[%
  pdftitle={Instructions for use of the document class
    elsart},%
  pdfauthor={Simon Pepping},%
  pdfsubject={The preprint document class elsart},%
  pdfkeywords={instructions for use, elsart, document class},%
  pdfstartview=FitH,%
  bookmarks=true,%
  bookmarksopen=true,%
  breaklinks=true,%
  colorlinks=true,%
  linkcolor=blue,anchorcolor=blue,%
  citecolor=blue,filecolor=blue,%
  menucolor=blue,pagecolor=blue,%
  urlcolor=blue]{hyperref}
\else
\usepackage[%
  breaklinks=true,%
  colorlinks=true,%
  linkcolor=blue,anchorcolor=blue,%
  citecolor=blue,filecolor=blue,%
  menucolor=blue,pagecolor=blue,%
  urlcolor=blue]{hyperref}
\fi

\renewcommand{\theequation}{\arabic{section}.\arabic{equation}}
\newtheorem{theorem}{Theorem}[section]
\newtheorem{remark}{Remark}[section]

\newenvironment{proof}[1][Proof]{\noindent\emph{{#1.}} }{ \rule{0.5em}{0.5em}}

\def\@subjclass{}
\makeatletter
\def\elsartstyle{%
    \def\normalsize{\@setfontsize\normalsize\@xiipt{14.5}}
    \def\small{\@setfontsize\small\@xipt{13.6}}
    \let\footnotesize=\small
    \def\large{\@setfontsize\large\@xivpt{18}}
    \def\Large{\@setfontsize\Large\@xviipt{22}}
    \skip\@mpfootins = 18\p@ \@plus 2\p@
    \normalsize
} \@ifundefined{square}{}{\let\Box\square} \makeatother
\usepackage{indentfirst}
\usepackage{setspace}
\begin{document}
\begin{frontmatter}
\title{Solitons, peakons, and periodic cuspons  of  a generalized
Degasperis-Procesi equation}
\author{Jiangbo Zhou\corauthref{cor}},
\corauth[cor]{Corresponding author.} \ead{zhoujiangbo@yahoo.cn}
\author{Lixin Tian}
\address{Nonlinear Scientific Research Center, Faculty of Science, Jiangsu
University, Zhenjiang, Jiangsu 212013, China}
\begin{abstract} In this paper, we employ the bifurcation theory of
planar dynamical systems to investigate the exact travelling wave
solutions of a generalized Degasperis-Procesi equation $
u_t-u_{xxt}+4u u_x+\gamma (u-u_{xx})_x=3u_xu_{xx}+uu_{xxx}$. The
implicit expression of smooth soliton solutions is given. The
explicit expressions of peaked soliton solutions and periodic cuspon
solutions are also obtained. Further, we show the relationship among
the smooth soliton solutions, the peaked solitons solution and the
periodic cuspon solutions. The physical relevance of the found
solutions and the reason why these solutions can exist in this
equation are also given.
\end{abstract}


\end{frontmatter}
\section{Introduction}
 \setcounter {equation}{0}
Recently, Degasperis and Procesi \cite{1} derived a nonlinear
dispersive equation
\begin{equation}
\label{eq1.1} u_t - u_{xxt} + 4uu_x = 3u_x u_{xx} + uu_{xxx}
\end{equation}
which is called the Degasperis-Procesi equation. Here $u(t,x)$
represents the fluid velocity at time $t$ in the $x$ direction in
appropriate nondimensional units (or, equivalently the height of the
water's free surface above a flat bottom). The nonlinear convection
term $uu_x$ in Eq.(\ref{eq1.1}) causes the steepening of wave form,
whereas the nonlinear dispersion effect term
$3u_xu_{xx}+uu_{xxx}=(\frac{1}{2}u^2)_{xxx}$ in Eq.(\ref{eq1.1})
makes the wave form spread. Eq.(\ref{eq1.1}) can be regarded as a
model for nonlinear shallow water dynamics \cite{2}. Degasperis,
Holm and Hone \cite{2} showed that the Eq.(\ref{eq1.1}) is
integrable by deriving a Lax pair and a bi-Hamiltonian structure for
it. Yin proved local well-posedness to Eq.(\ref{eq1.1}) with initial
data $u_0 \in H^s(\mathbb{R})$, $s > 3$ on the line \cite{3} and on
the circle \cite{4}. The global existence of strong solutions and
weak solutions to Eq.(\ref{eq1.1}) are investigated in
\cite{4}-\cite{10}. The solution to Cauchy problem of
Eq.(\ref{eq1.1}) can also blow up in finite time when the initial
data satisfies certain sign condition\cite{7}-\cite{10}. Vakhnenko
and Parkes \cite{11} obtained periodic and solitary-wave solutions
of Eq.(\ref{eq1.1}). Matsuno \cite{12, 13} obtained multisoliton,
cusp and loop soliton solutions of Eq.(\ref{eq1.1}). Lundmark and
Szmigielski \cite{14} investigated multi-peakon solutions of
Eq.(\ref{eq1.1}).  Lenells \cite{15} classified all weak traveling
wave solutions. The shock wave solutions of Eq.(\ref{eq1.1}) are
investigated in \cite{16, 17}.

Yu and Tian \cite {18} investigated the following generalized
Degasperis-Procesi equation
\begin{equation}
 \label {eq1.2} u_t - u_{xxt} + 4uu_x =3u_x
u_{xx}+uu_{xxx}-\gamma u_{xxx},
\end{equation}
where $\gamma$ is a real constant, and the term $ u_{xxx}$ denotes
the linear dispersive effect. They obtained peaked soliton solutions
and period cuspon solutions of Eq.(\ref{eq1.2}). Unfortunately, they
didn't obtain smooth soliton solutions of Eq.(\ref{eq1.2}).

In this paper, we are interesting in the following generalized
Degasperis-Procesi equation
\begin{equation}
\label{eq1.3}
 u_t-u_{xxt}+4u u_x+\gamma (u-u_{xx})_x=3u_xu_{xx}+uu_{xxx},
\end{equation}
where $\gamma$ is a real constant, the term $u_x$  denotes the
dissipative effect and the term $ u_{xxx}$ represents the linear
dispersive effect. Employing the bifurcation theory of planar
dynamical systems, we obtain the analytic expressions of smooth
solitons, peaked solitons and period cuspons of Eq.(\ref{eq1.3}).
Our work covers and supplements the results obtained in \cite {18}.

The remainder of the paper is organized as follows. In Section 2,
using the travelling wave transformation,  we transform
Eq.(\ref{eq1.3}) into the planar dynamical system (\ref{eq2.3}) and
then discuss bifurcations of phase portraits of system
(\ref{eq2.3}). In Section 3, we obtain the implicit expression of
smooth solitons and the explicit expressions of peaked solitons and
periodic cuspon solutions. At the same time, we show that the limits
of smooth solitons and periodic cusp waves are peaked solitons. In
Section 4, we discuss the physical relevance of the found solutions
and give the reason why these solutions can exist in
Eq.(\ref{eq1.3}).

\section{Bifurcations of phase portraits of system (\ref{eq2.3})}
 \setcounter {equation}{0}
We look for travelling wave solutions of Eq.(\ref{eq1.3}) in the
form of $u(x,t) =\varphi (x-ct)= \varphi (\xi )$, where $c$ is the
wave speed and $\xi = x - ct$. Substituting  $u= \varphi (\xi )$
into Eq.(\ref{eq1.3}),  we obtain
\begin{equation}
\label{eq2.1}
 - c\varphi' + c\varphi ''' + 4\varphi \varphi ' -3\varphi' \varphi ''- \varphi \varphi ''' +\gamma \varphi ' -\gamma \varphi
 '''=0.
\end{equation}

Integrating (\ref{eq2.1}) once we have
\begin{equation}
\label{eq2.2} \varphi ''(\varphi - c+\gamma) = g -( c-\gamma)
\varphi + 2 \varphi ^2 - (\varphi ')^2,
\end{equation}
\noindent where $g$ is the integral constant.

Let $y = \varphi '$, then we get the following planar dynamical
system:
\begin{equation}
\label{eq2.3}\left\{ {\begin{array}{l}
 \frac{\textstyle  d\varphi }{\textstyle  d\xi } = y \\
 \frac{\textstyle dy}{\textstyle d\xi } = \frac{\textstyle g -( c-\gamma)
\varphi + 2 \varphi ^2 - y^2}{\textstyle  \varphi - c+\gamma} \\
 \end{array}} \right.
\end{equation}
\noindent with a first integral
\begin{equation}
\label{eq2.4} H(\varphi, y)=(\varphi - c+\gamma)^2\left(y^2 -
\varphi^2- g \right)=h,
\end{equation}
\noindent where $h$ is a constant.

Note that (\ref{eq2.3}) has a singular line $\varphi = c-\gamma$. To
avoid the line temporarily we make transformation $d\xi = (\varphi -
c+\gamma)d\zeta $. Under this transformation, Eq.(\ref{eq2.3})
becomes
\begin{equation}
\label{eq2.5} \left\{ {\begin{array}{l}
 \frac{\textstyle d\varphi }{\textstyle d\zeta } = (\varphi - c+\gamma)y \\
 \frac{\textstyle dy}{\textstyle d\zeta } = g -( c-\gamma)
\varphi + 2 \varphi ^2 - y^2
\\
 \end{array}} \right.
\end{equation}

System (\ref{eq2.3}) and system (\ref{eq2.5}) have the same first
integral as (\ref{eq2.4}). Consequently, system (\ref{eq2.5}) has
the same topological phase portraits as system (\ref{eq2.3}) except
for the straight line $\varphi = c-\gamma$. Obviously, $\varphi =
c-\gamma$ is an invariant straight-line solution for system
(\ref{eq2.5}).

 For a fixed $h$, (\ref{eq2.4}) determines a
set of invariant curves of system (\ref{eq2.5}). As $h$ is varied,
(\ref{eq2.4}) determines different families of orbits of system
(\ref{eq2.5}) having different dynamical behaviors. Let $M(\varphi
_e ,y_e )$ be the coefficient matrix of the linearized system of
(\ref{eq2.5}) at the equilibrium point $(\varphi _e ,y_e )$, then
\begin{equation}
\label{eq2.6} M(\varphi _e ,y_e ) = \left( {{\begin{array}{*{20}c}
{\indent  y_e } \hfill &&& {\varphi _e - c+\gamma} \hfill \\
 {4\varphi _e - c +\gamma} \hfill &&& \indent{- 2y_e} \hfill \\
\end{array} }} \right)
\end{equation}
and at this equilibrium point, we have
\begin{equation}
\label{eq2.7} J(\varphi _e ,y_e ) = \det M(\varphi _e ,y_e ) = -
2y_e^2 - (\varphi _e - c+\gamma)(4\varphi _e - c +\gamma),
\end{equation}
\begin{equation}
\label{eq2.8} p(\varphi _e ,y_e ) = \mathrm{trace}(M(\varphi _e ,y_e
)) = - y_e.
\end{equation}
By the qualitative theory of differential equations (see \cite
{19}), for an equilibrium point of a planar dynamical system, if $J
< 0$, then this equilibrium point is a saddle point; it is a center
point if $J
> 0$ and $p = 0$; if $J = 0$ and the Poincar\'{e} index of the
equilibrium point is 0, then it is a cusp.

By using the first integral value and properties of equilibrium
points, we obtain the bifurcation curves as follows:
\begin{equation}
\label{eq2.9} g_1 (c) = \frac{(c - \gamma)^2}{8},
\end{equation}

\begin{equation}
\label{eq2.10} g_2 (c) = -(c - \gamma)^2.
\end{equation}
Obviously, the two curves have no intersection point and
$g_2(c)<0<g_1(c)$ for arbitrary constants $c\neq\gamma$.

Using bifurcation method of vector fields (e.g., \cite{19}), we have
the following result which describes the locations and properties of
the singular points of system (\ref{eq2.5}).
\begin{theorem}
For given any constant wave speed $c\neq 0$, let
\begin{equation}
\label{eq2.11}\varphi _0^\pm =\frac { c - \gamma \pm \sqrt {(c -
\gamma)^2 - 8g}}{4}\quad for \quad g\leq g_1(c),
\end{equation}
\begin{equation}
\label{eq2.12} y _0^\pm =\pm \sqrt {(c - \gamma)^2 +g }\quad for
\quad g\geq g_2(c).
\end{equation}

When $c=\gamma$,

(1) if $g<0$, then system (\ref{eq2.5}) has two equilibrium points
$(-\sqrt{-\frac{g}{2}} ,0)$ and $(\sqrt{-\frac{g}{2}} ,0)$, which
are saddle points.

(2) if $g=0 $, then system (\ref{eq2.5}) has only one equilibrium
point $(0,0)$, which is a cusp.

(3) if $g>0$, then system (\ref{eq2.5}) has two equilibrium points
$(0 ,-\sqrt{g})$ and $(0 ,\sqrt{g})$, which are saddle points.

When $c\neq\gamma$,

(1) if $g < g_2 (c)$, then system (\ref{eq2.5}) has two equilibrium
points $(\varphi _0^ - ,0)$ and $(\varphi _0^ + ,0)$. They are
saddle points.

(i) if $c> \gamma$, then $\varphi _0^ - <
-\frac{1}{2}(c-\gamma)<\frac{1}{4}(c-\gamma)< c-\gamma < \varphi
_0^+$.

(ii) if $c< \gamma$, then $\varphi _0^ - <c-\gamma
<\frac{1}{4}(c-\gamma) <-\frac{1}{2}(c-\gamma) < \varphi _0^+$.

(2) if $ g=g_2 (c)$, then system (\ref{eq2.5}) has three equilibrium
points $(\varphi _0^ - ,0)$,  $(\varphi _0^ + ,0)$ and $(c-\gamma
,0)$. $(c-\gamma ,0)$ is a cusp.

(i) if $c> \gamma$, then $\varphi _0^ - =
-\frac{1}{2}(c-\gamma)<\frac{1}{4}(c-\gamma)< c-\gamma = \varphi
_0^+$.  $(\varphi _0^ - ,0)$ is a saddle point, while  $(\varphi _0^
+ ,0)$ is a degenerate center point.

(ii) if  $c< \gamma$, then $\varphi _0^ - =c-\gamma
<\frac{1}{4}(c-\gamma) <-\frac{1}{2}(c-\gamma) = \varphi _0^+$.
$(\varphi _0^ - ,0)$ is a degenerate center point, while $(\varphi
_0^ + ,0)$ is a saddle point.

(3) if  $g_2 (c) < g < g_1 (c)$, then system (\ref{eq2.5}) has four
equilibrium points  $(\varphi _0^ - ,0)$,  $(\varphi _0^ + ,0)$,
$(c-\gamma,y_0^-)$ and $(c-\gamma,y_0^+)$. $(c-\gamma,y_0^-)$ and
$(c-\gamma,y_0^+)$ are two saddle points.

(i)  if $c> \gamma$, then $\varphi _0^ - <
\frac{1}{4}(c-\gamma)<\varphi _0^+\leq c-\gamma $. $(\varphi _0^ -
,0)$ is a saddle point, while  $(\varphi _0^ + ,0)$ is a center
point.

(ii) if  $c<\gamma$, then $c-\gamma\leq \varphi
_0^-<\frac{1}{4}(c-\gamma)< \varphi _0^+$.  $(\varphi _0^ - ,0)$ is
a center point, while  $(\varphi _0^ + ,0)$ is a saddle point.

Specially, when  $ g =0$,

(i) if $c> \gamma$, then the three saddle points $(\varphi _0^ -
,0)$, $(c-\gamma,y_0^-)$ and $(c-\gamma,y_0^+)$ form a triangular
orbit which encloses the center point $(\varphi _0^ + ,0)$.

(ii) if $c<\gamma$, then the three saddle points $(\varphi _0^ +
,0)$, $(c-\gamma,y_0^-)$ and $(c-\gamma,y_0^+)$ form a triangular
orbit which encloses the center point $(\varphi _0^ - ,0)$.

(4) if $g=g_1 (c)$, then system (\ref{eq2.5}) has three equilibrium
points  $(\frac{c-\gamma}{4},0)$, $(c-\gamma,y_0^-)$ and
$(c-\gamma,y_0^+)$.
 $(\frac{c-\gamma}{4},0)$ is a degenerate center point, while $(c-\gamma,y_0^-)$ and $(c-\gamma,y_0^+)$
are two saddle points.

(5) if $g>g_1(c)$, then system (\ref{eq2.5}) has two equilibrium
points $(c-\gamma,y_0^-)$ and $(c-\gamma,y_0^+)$, which are saddle
points.

\end{theorem}

\begin{figure}[h]
\centering
\subfloat[]{\includegraphics[height=1.4in,width=1.5in]{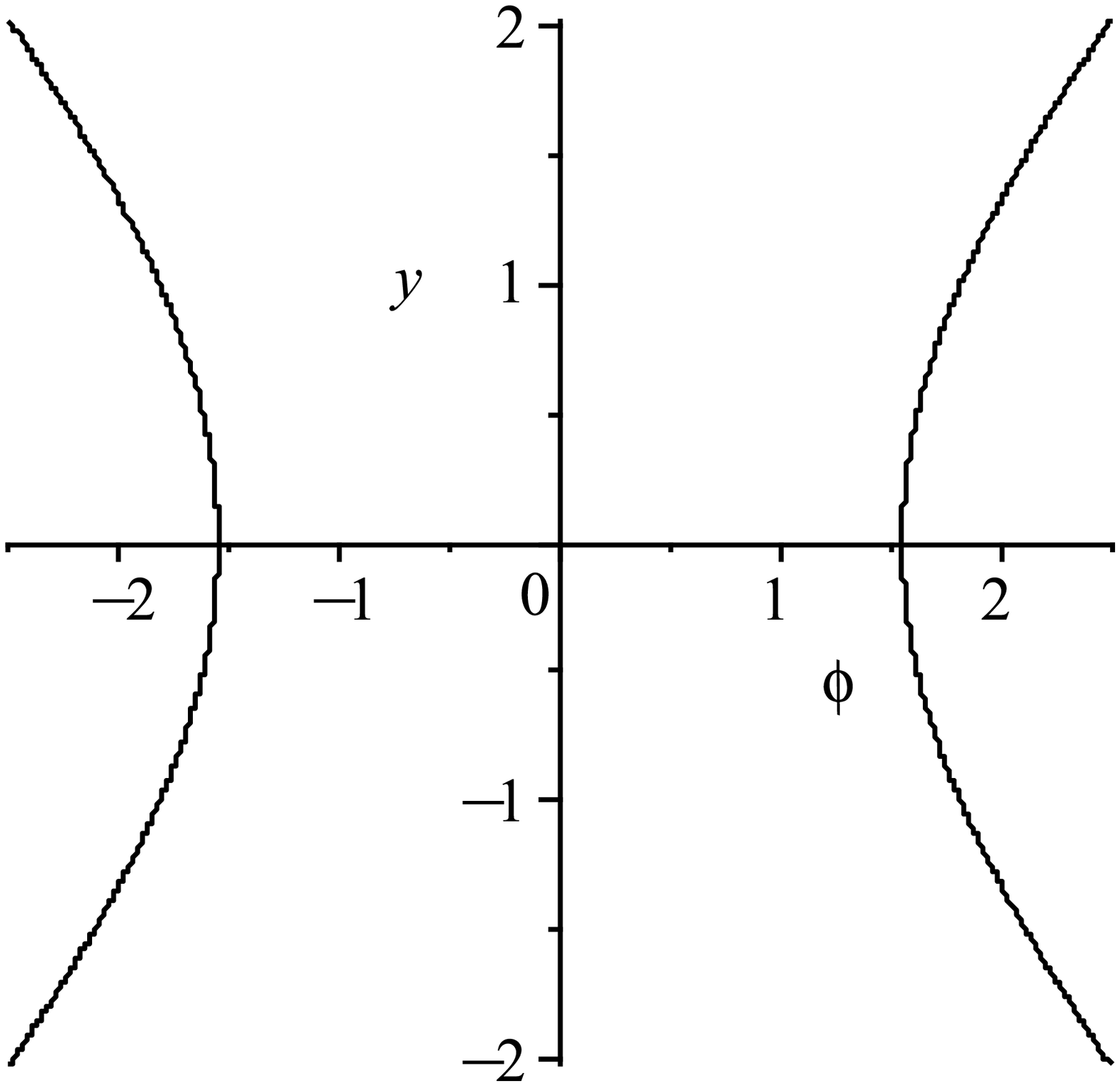}}\hspace{0.05\linewidth}
\subfloat[]{\includegraphics[height=1.4in,width=1.5in]{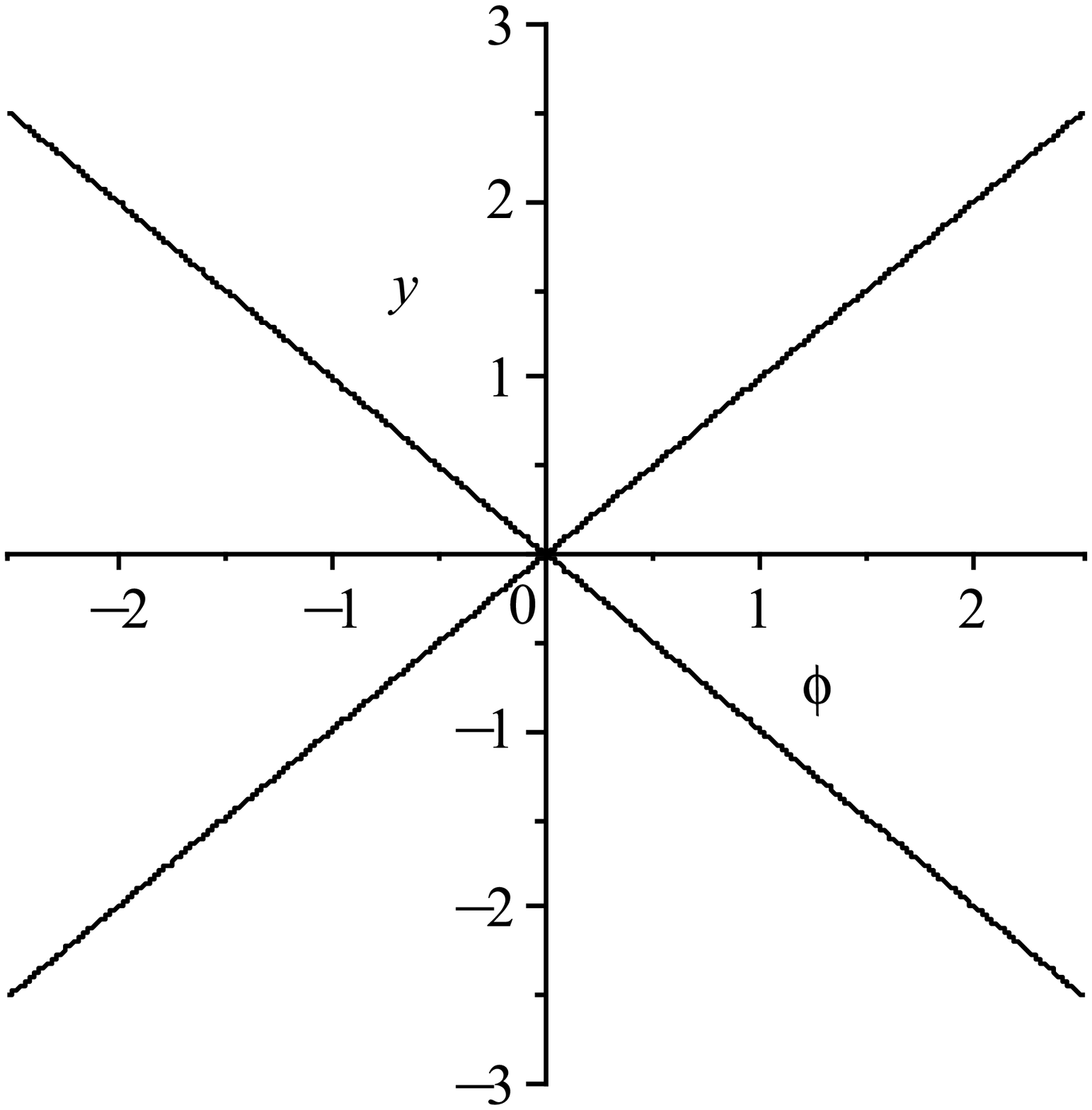}}\hspace{0.05\linewidth}
\subfloat[]{\includegraphics[height=1.4in,width=1.5in]{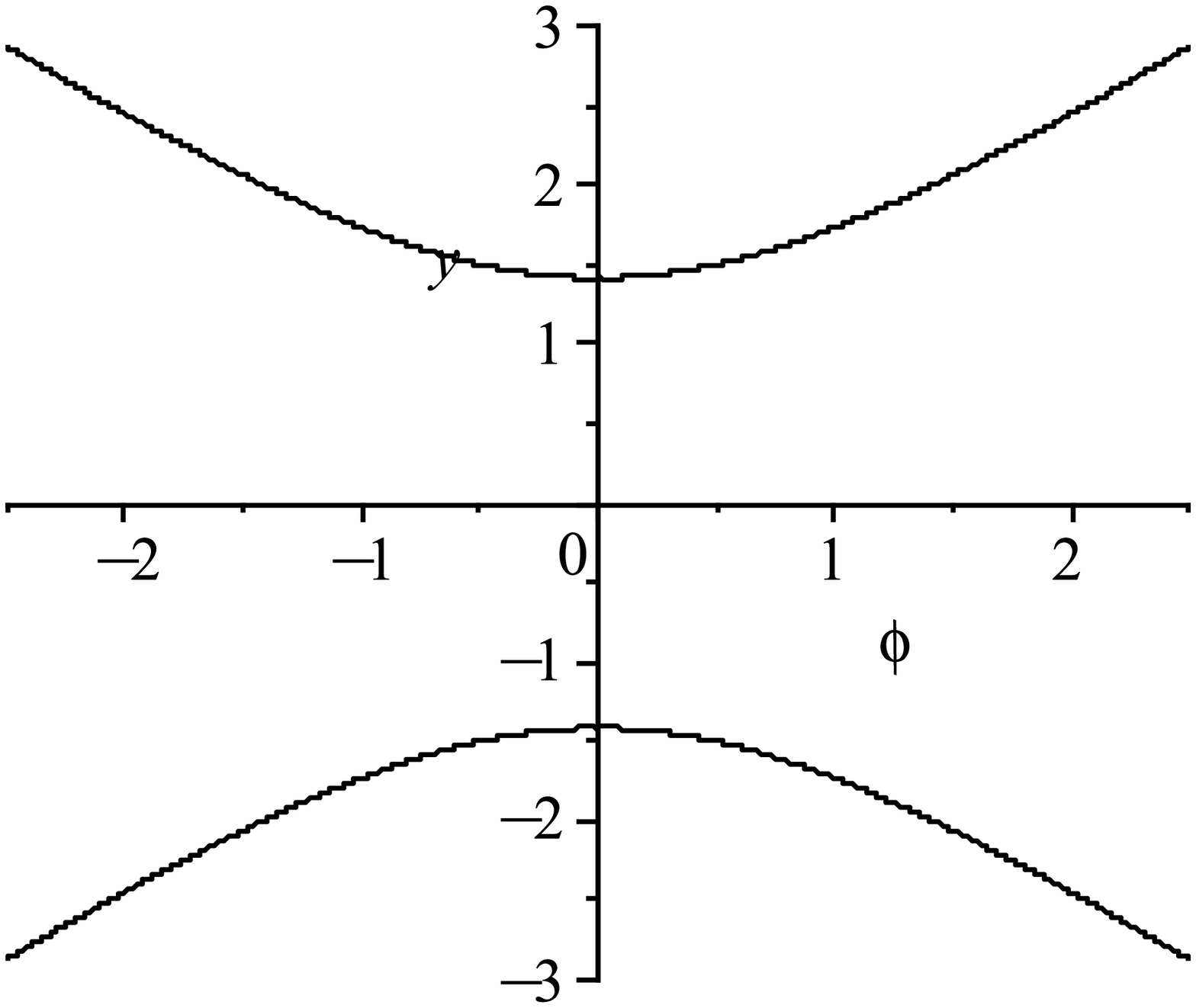}}\\
 \caption {The phase portraits of system
(\ref{eq2.5})($c=\gamma$). (a) $g<0$; (b) $g=0$; (c) $g>0$.
}\label{f1}
\end{figure}

\begin{figure}[h]
\centering
\subfloat[]{\includegraphics[height=1.3in,width=1.4in]{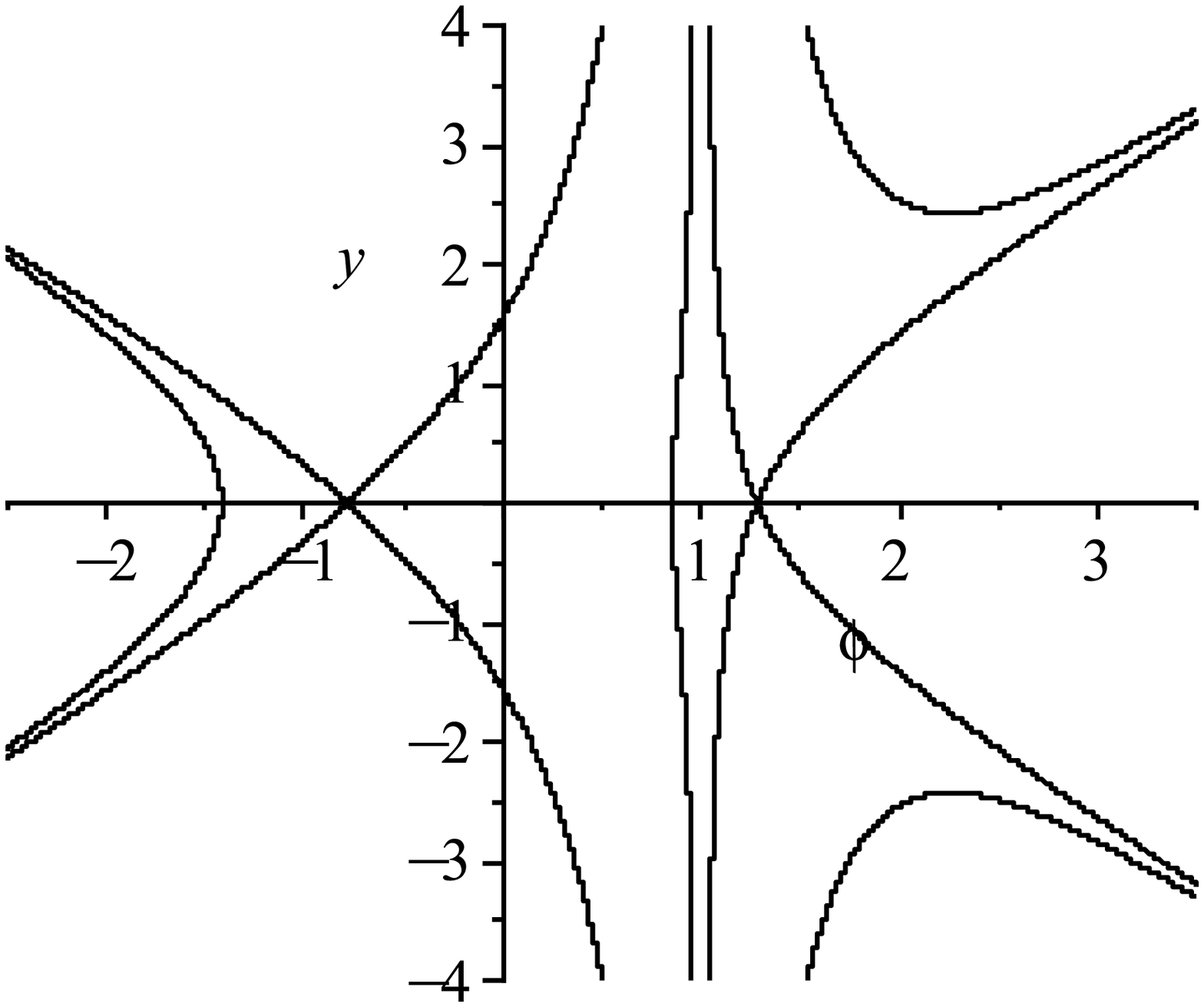}}
\subfloat[]{\includegraphics[height=1.3in,width=1.4in]{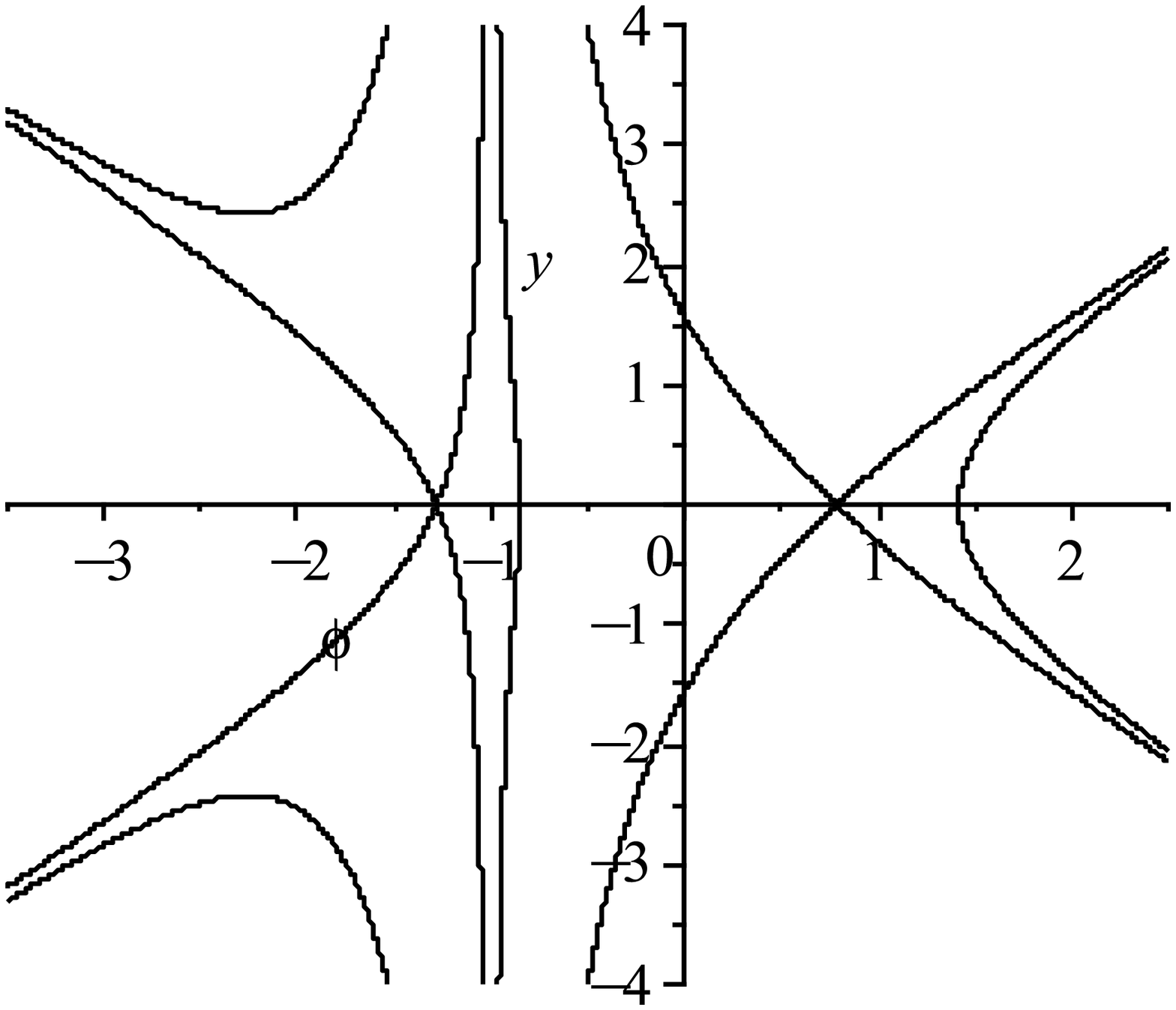}}
\subfloat[]{\includegraphics[height=1.3in,width=1.4in]{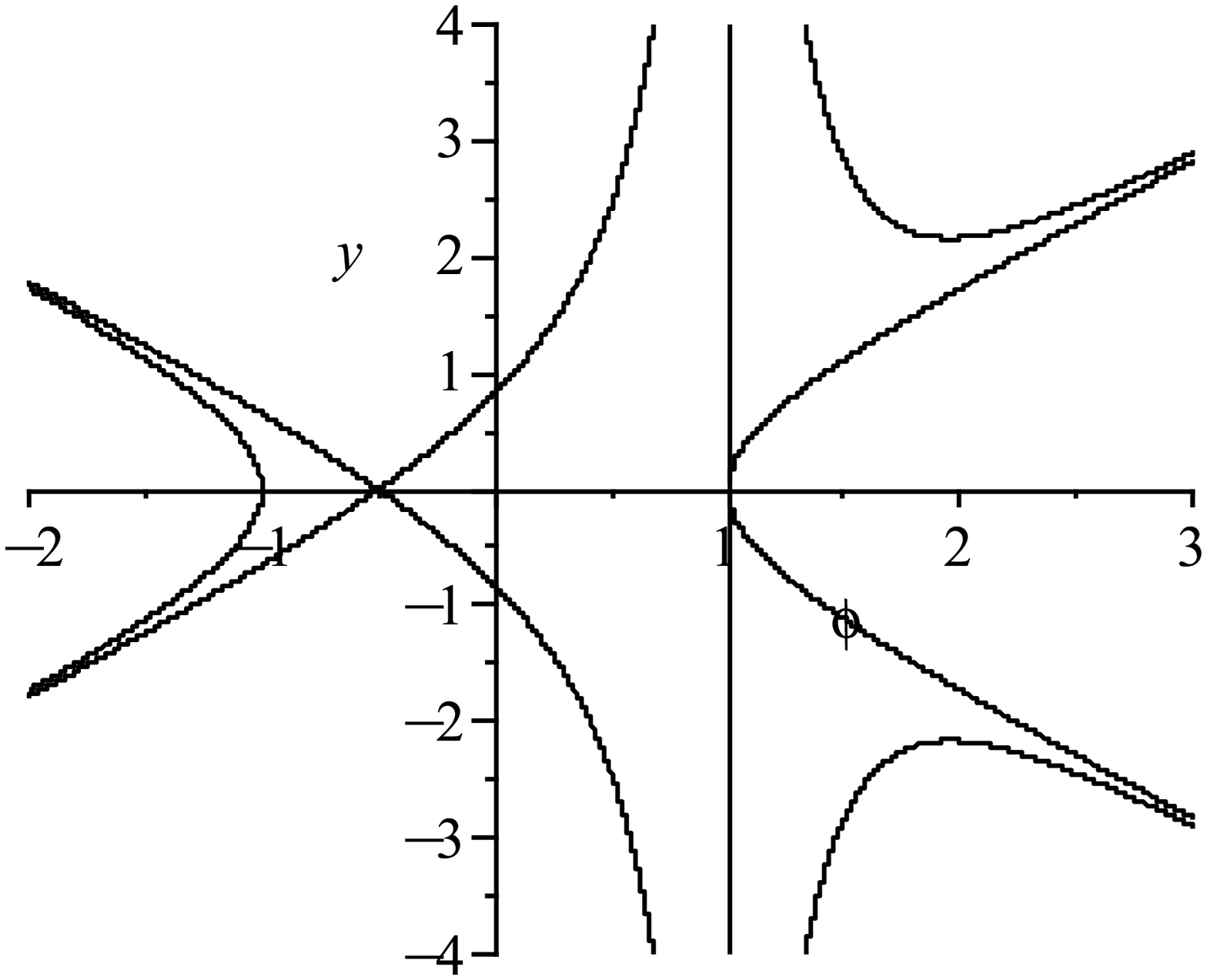}}
\subfloat[]{\includegraphics[height=1.3in,width=1.4in]{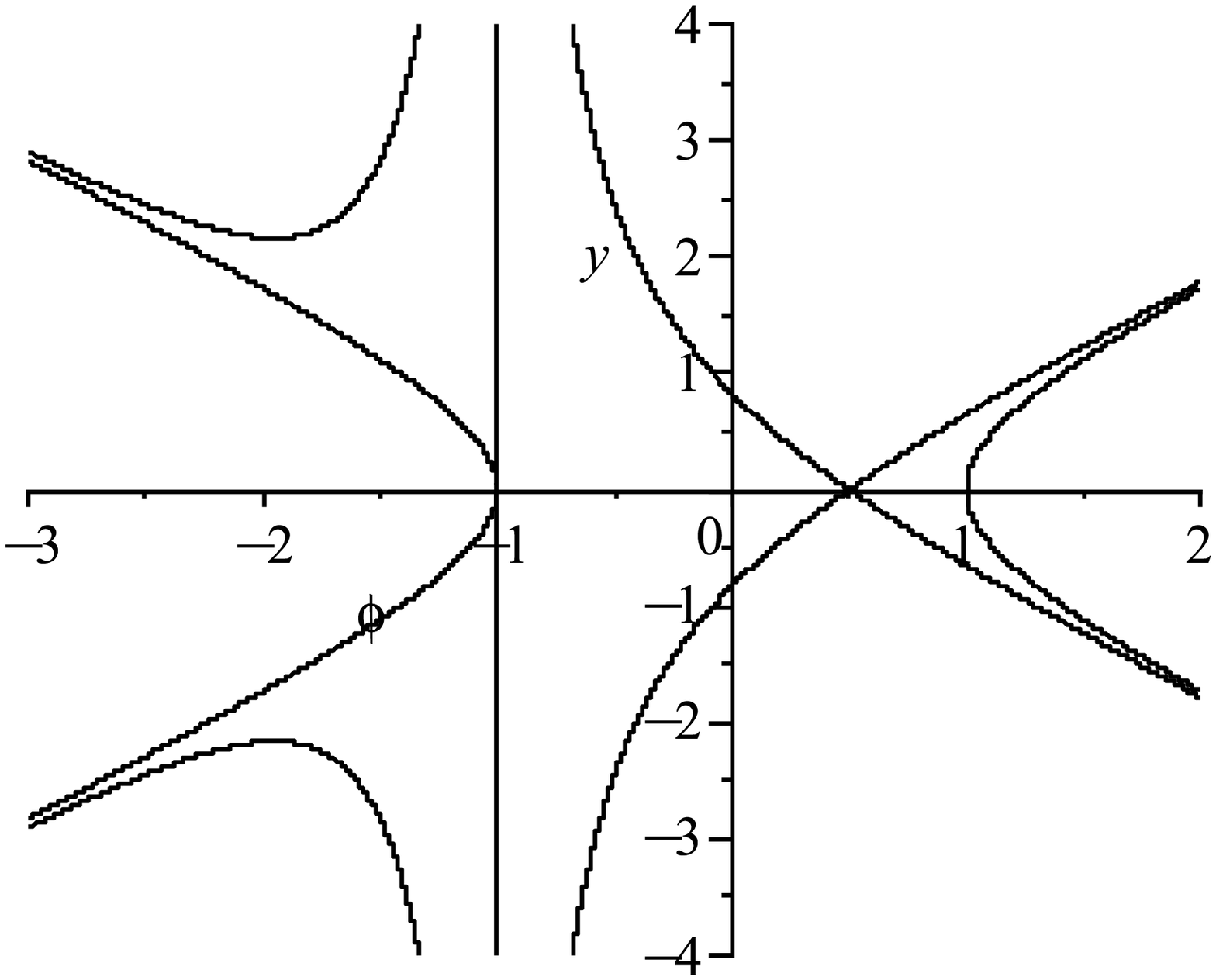}}\\
\subfloat[]{\includegraphics[height=1.3in,width=1.4in]{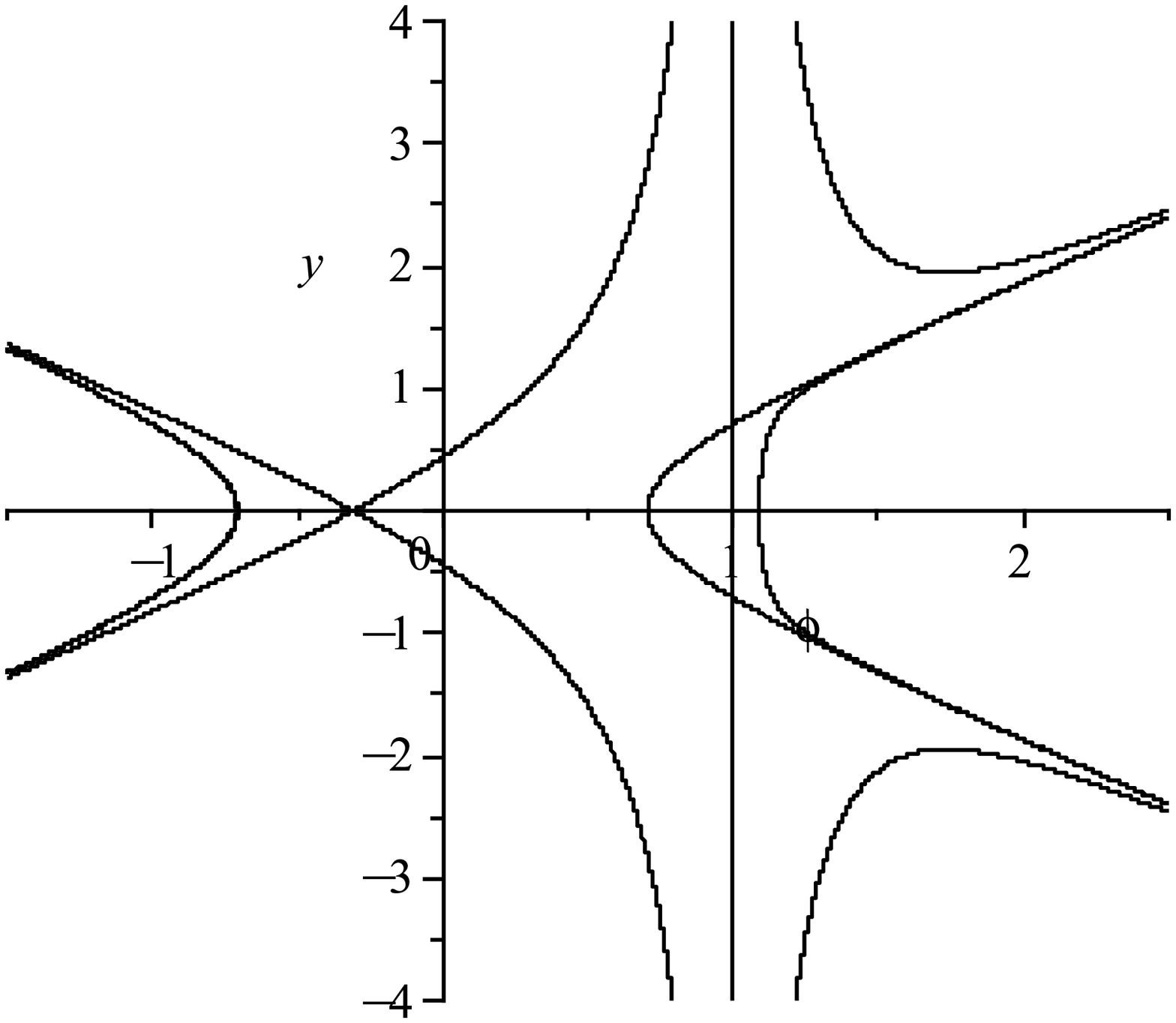}}
\subfloat[]{\includegraphics[height=1.3in,width=1.4in]{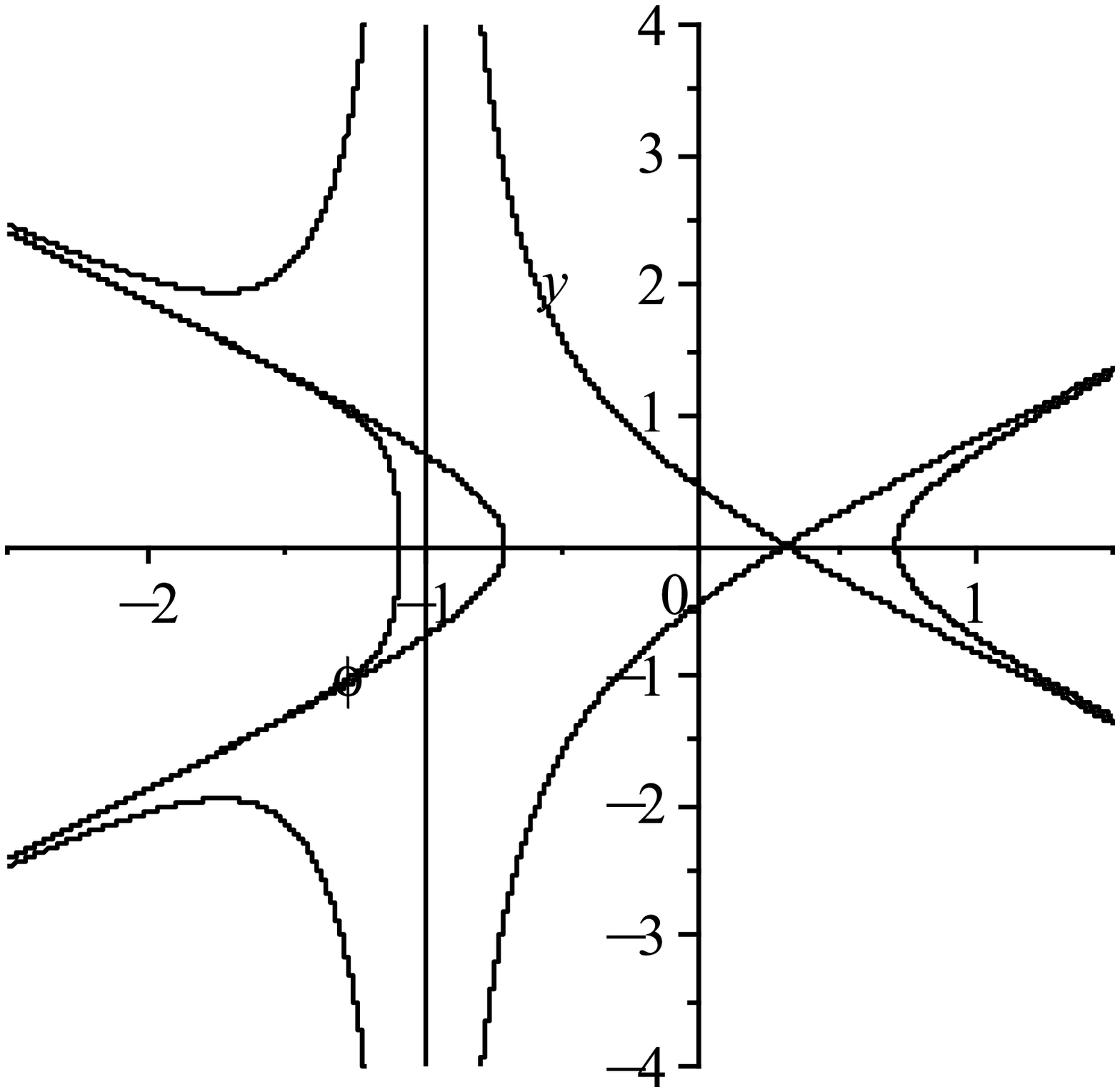}}
\subfloat[]{\includegraphics[height=1.3in,width=1.4in]{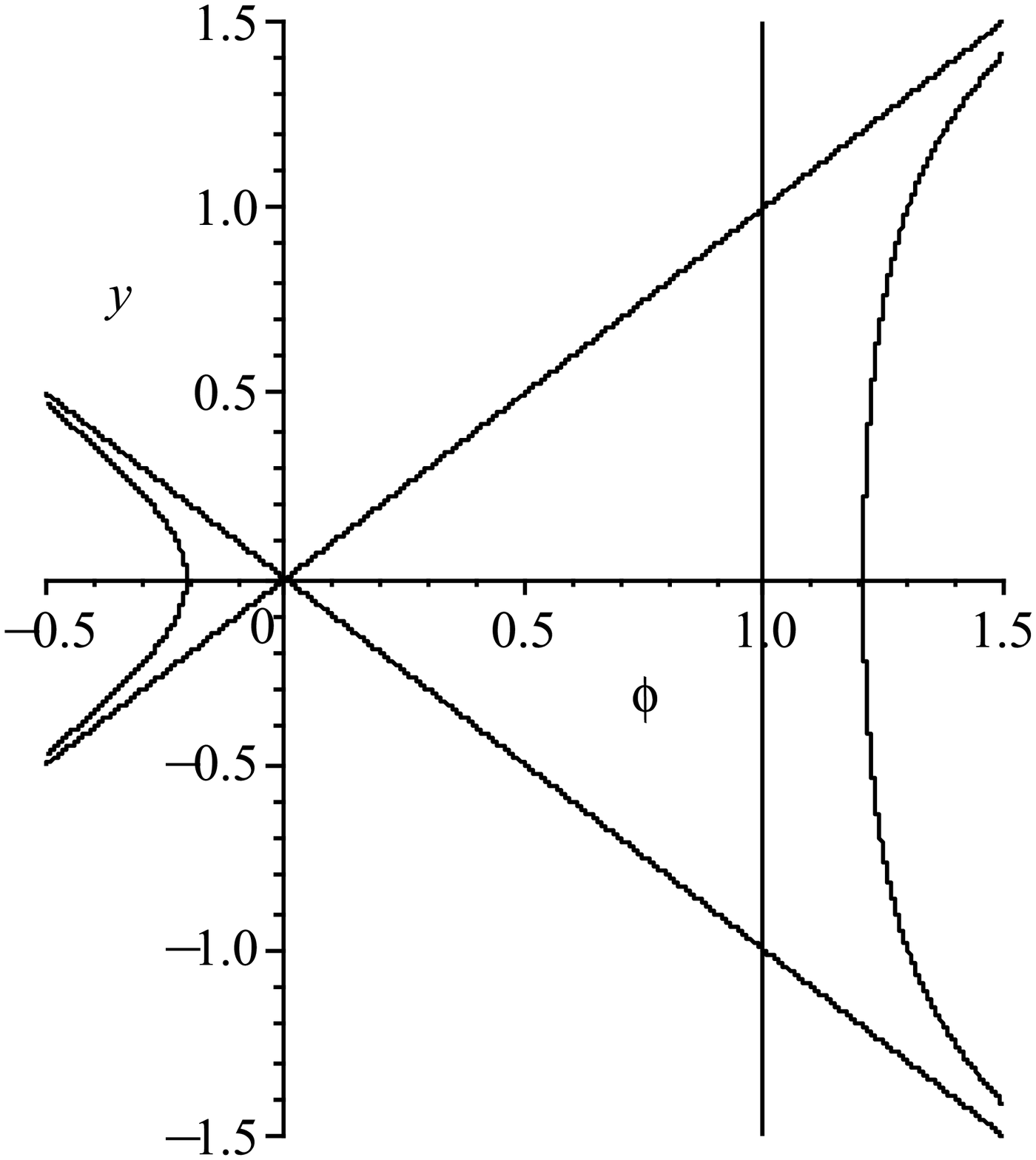}}
\subfloat[]{\includegraphics[height=1.3in,width=1.4in]{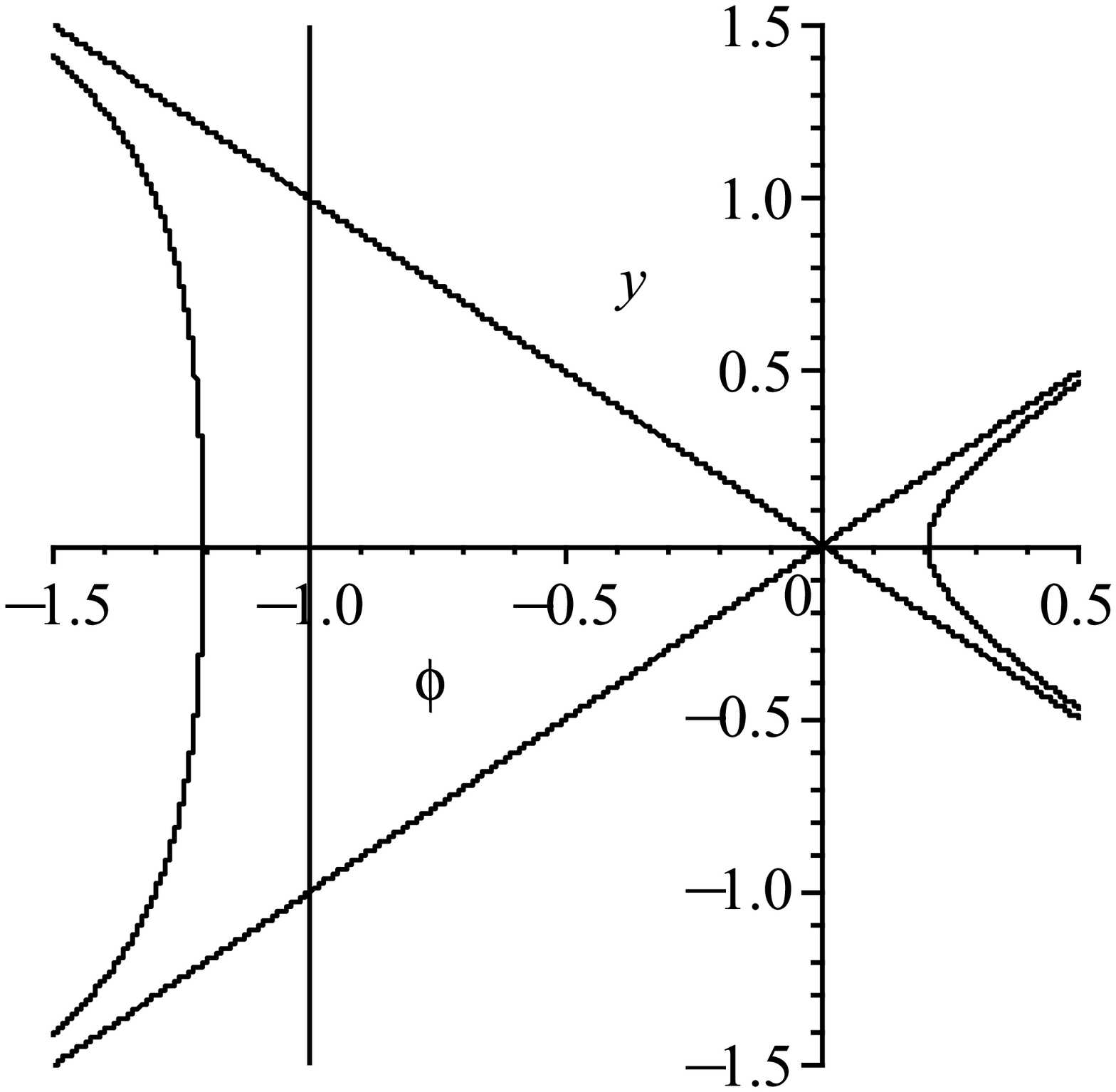}}\\
\subfloat[]{\includegraphics[height=1.3in,width=1.4in]{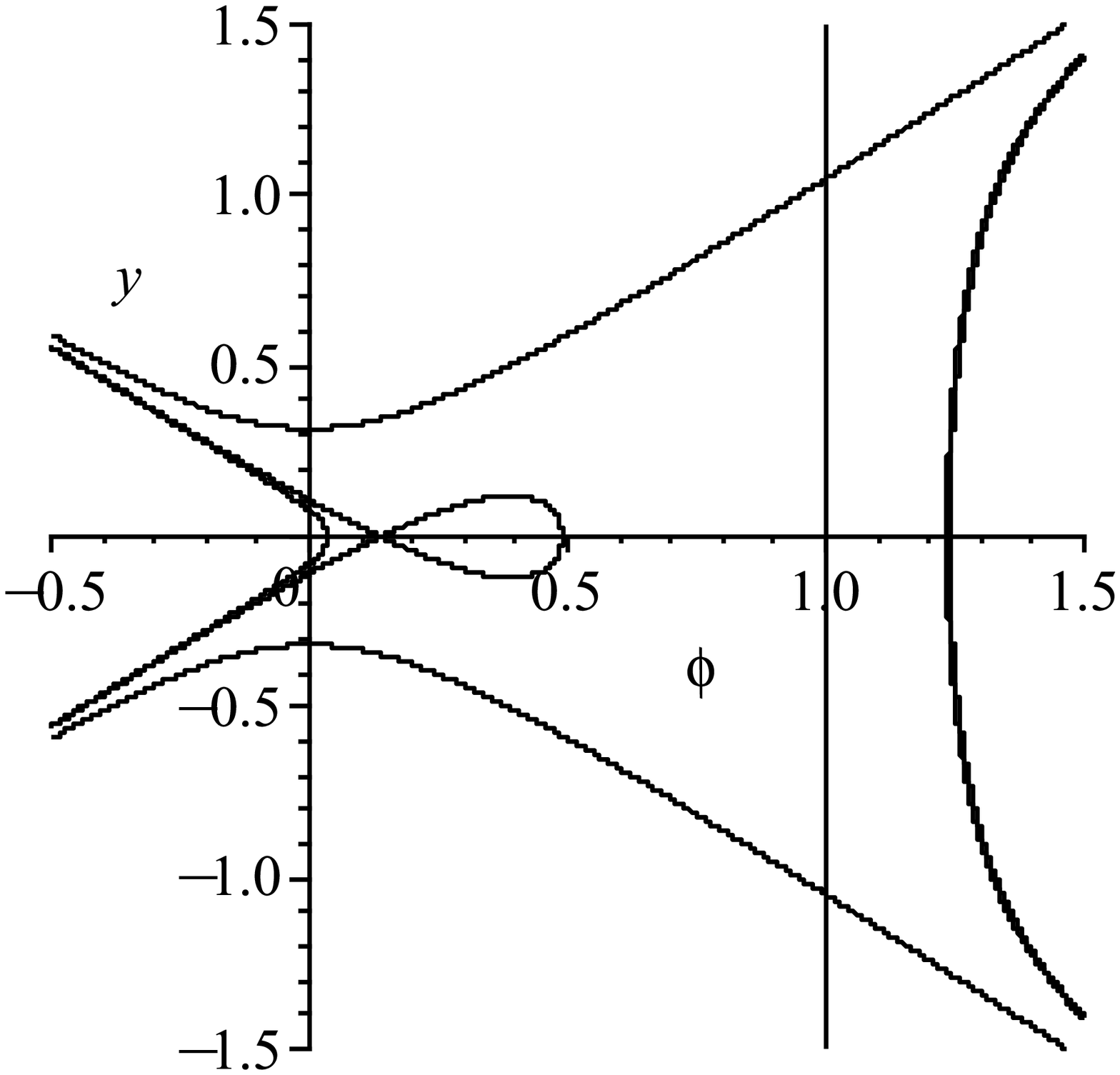}}
\subfloat[]{\includegraphics[height=1.3in,width=1.4in]{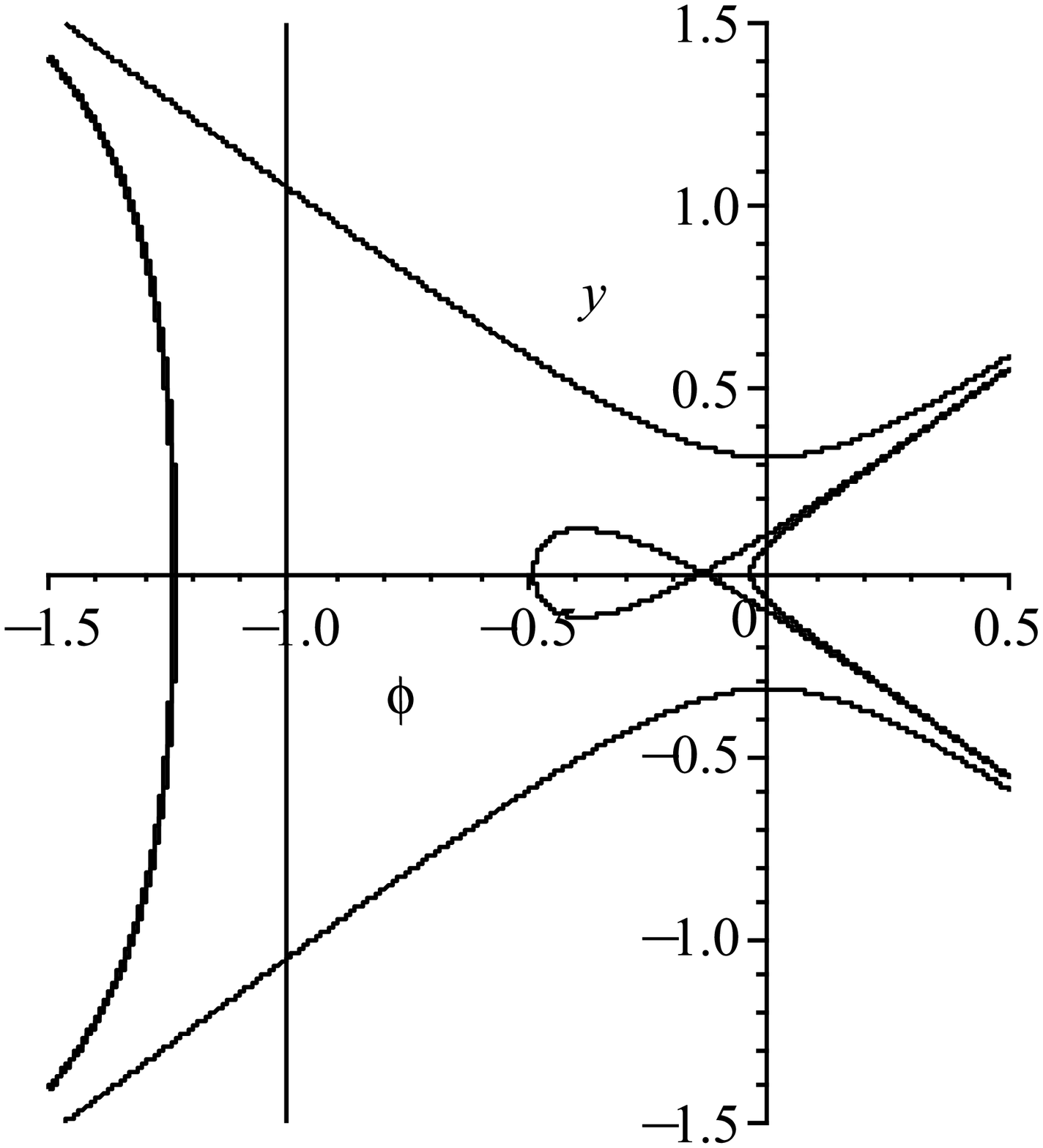}}
\subfloat[]{\includegraphics[height=1.3in,width=1.4in]{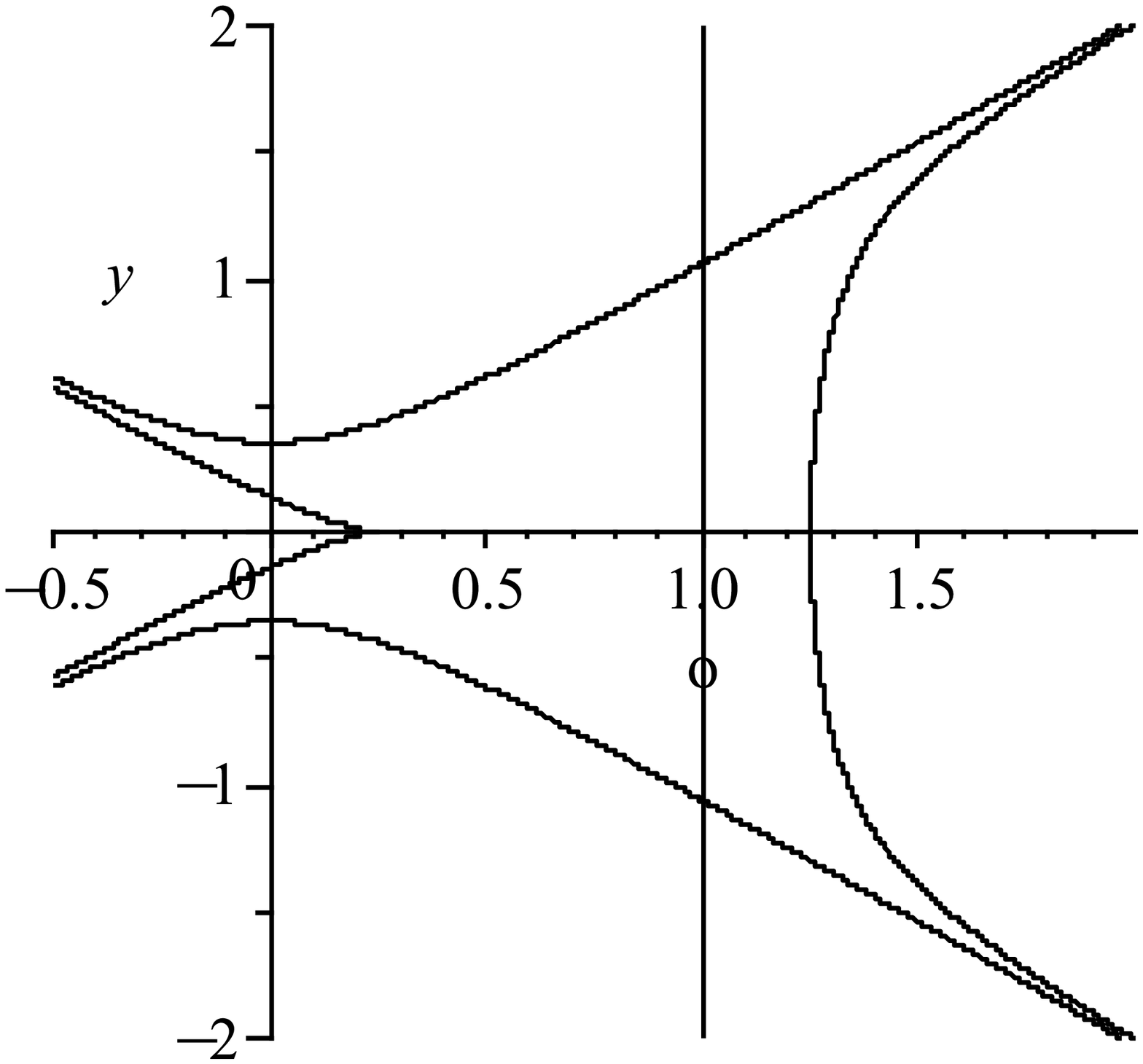}}
\subfloat[]{\includegraphics[height=1.3in,width=1.4in]{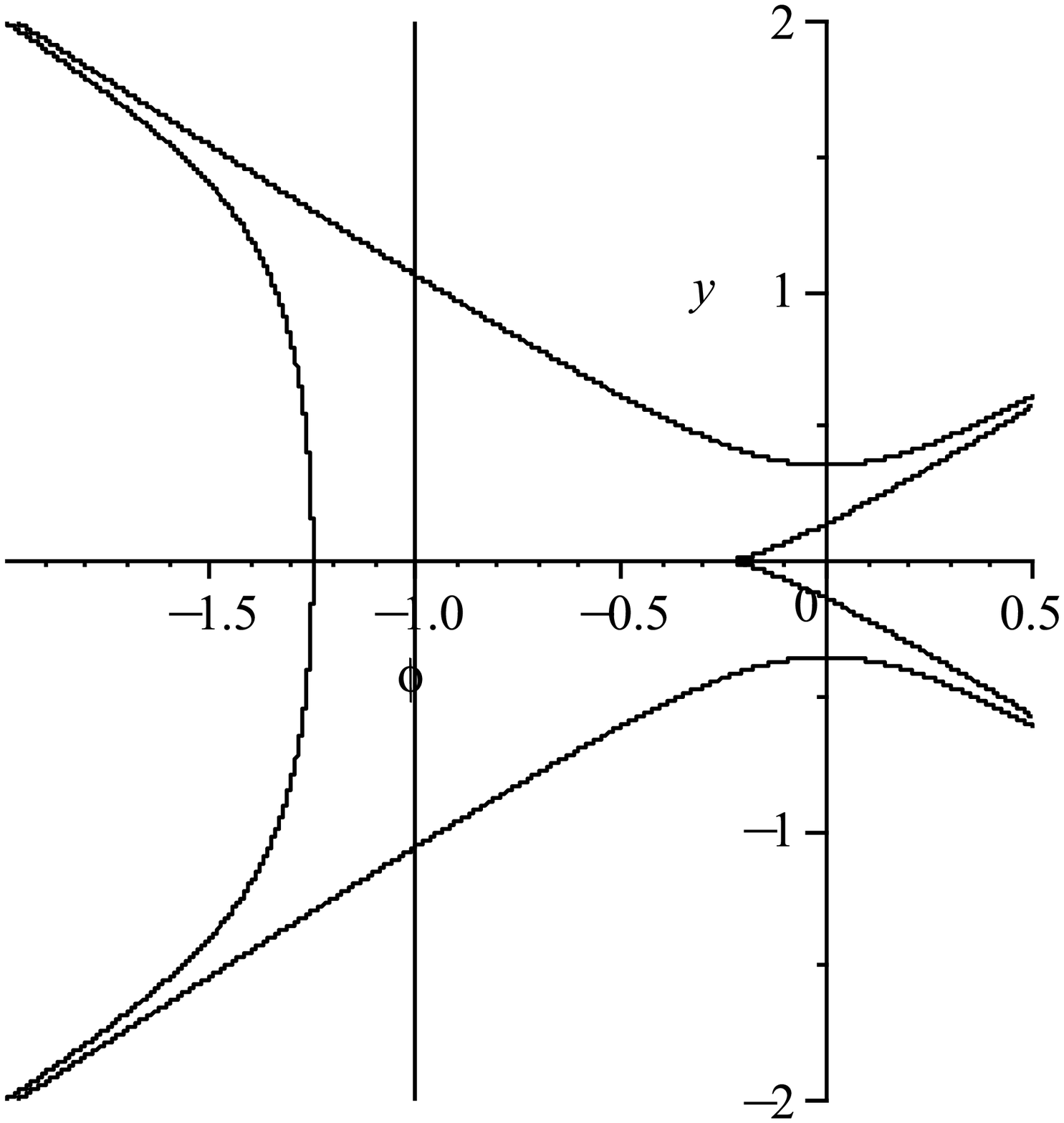}}\\
\subfloat[]{\includegraphics[height=1.4in,width=1.5in]{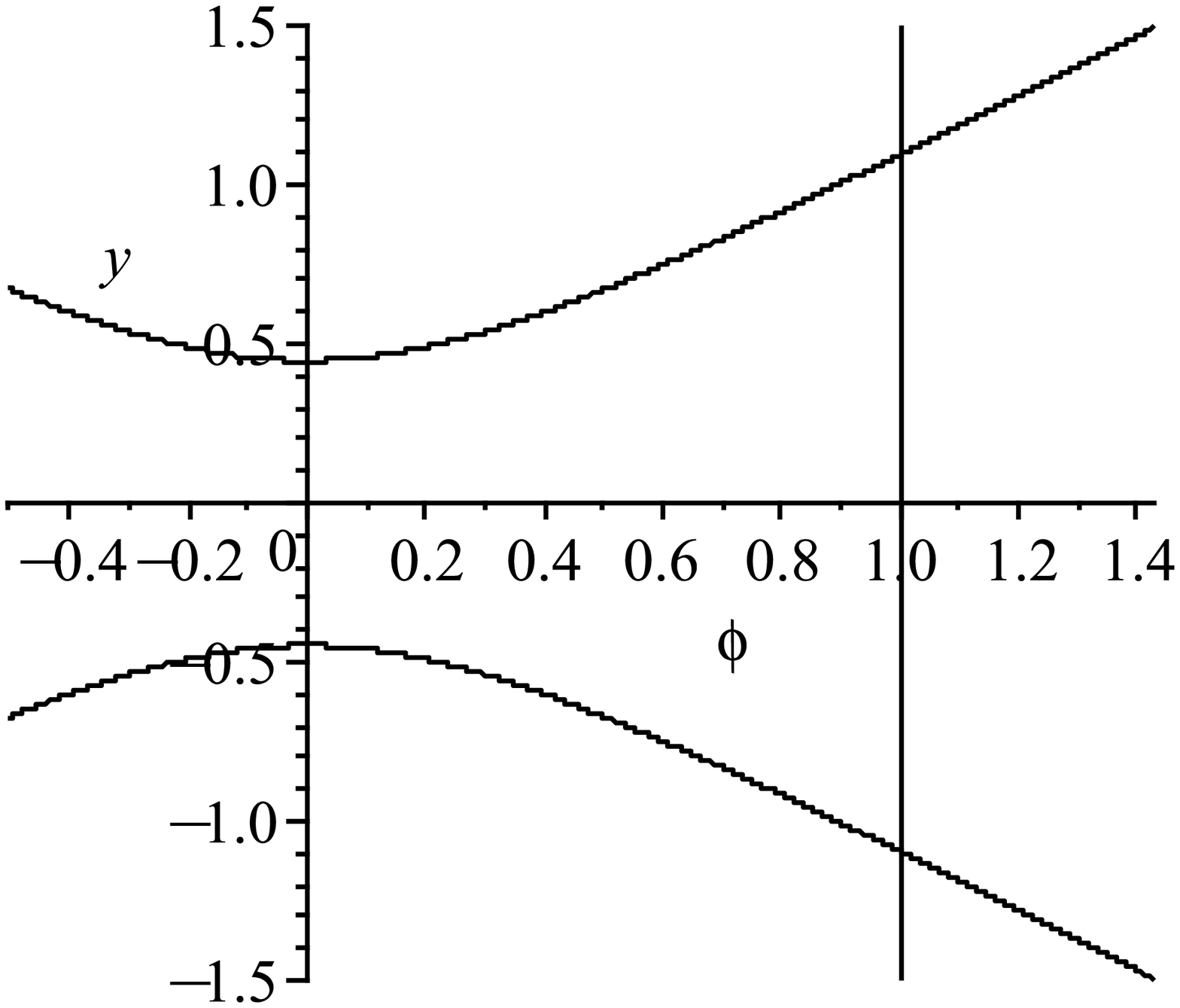}}\hspace{0.15\linewidth}
\subfloat[]{\includegraphics[height=1.4in,width=1.5in]{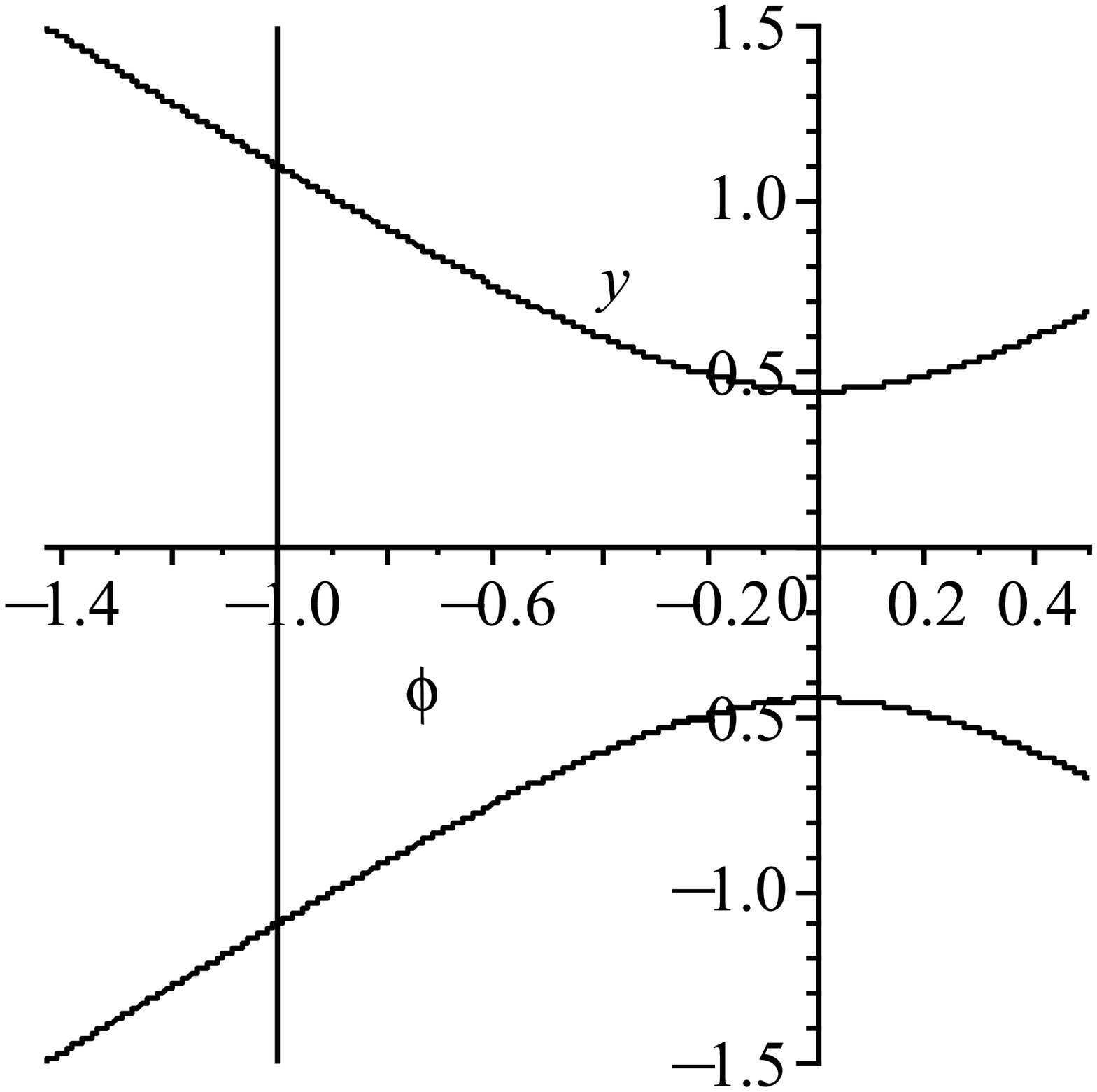}}
\caption { The phase portraits of system (\ref{eq2.5})
($c\neq\gamma$). (a) $g<g_2(c), c>\gamma$; (b) $g<g_2(c),c<\gamma$;
(c) $g=g_2(c), c>\gamma$; (d)$g=g_2(c), c<\gamma$; (e)  $g_2(c)<g<0,
c>\gamma$;  (f) $g_2(c)<g<0, c<\gamma$; (g) $g=0,c>\gamma $;
 (h) $g=0, c<\gamma $; (i) $0<g<g_1(c), c>\gamma$; (j)$0<g<g_1(c), c<\gamma$; (k) $g=g_1(c), c>\gamma$;
 (l) $g=g_1(c), c<\gamma$;(m)$g>g_1(c), c>\gamma$;(n)$g>g_1(c),
 c<\gamma$.}\label{f2}
\end{figure}

Corresponding to the case $c=\gamma$ and the case $c\neq\gamma$, we
show the phase portraits of system (\ref{eq2.5}) in Fig.\ref{f1} and
Fig.\ref{f2}, respectively.

\section{Solitons, peakons and periodic cusp wave solutions}
 \setcounter {equation}{0}
\begin{theorem}
\label{th2} Given arbitrary constant $c\neq\gamma$, let $ \xi = x -
ct$, then

(1) when $0<g < g_1 (c)$,

(i) if $ c>\gamma$, then Eq.(\ref{eq1.3}) has the following smooth
hump-like soliton solutions
\begin{equation}
\label{eq3.1} \beta_1(\varphi _1^+) = \beta_1(\varphi) e^{ - |\xi| }
\quad for\quad \varphi_0^-<\varphi<\varphi _1^+,
\end{equation}

(ii) if $ c<\gamma$, then Eq.(\ref{eq1.3}) has the following smooth
valley-like soliton solutions
\begin{equation}
\label{eq3.2} \beta_2(\varphi ) = \beta_2(\varphi_1^-) e^{ - |\xi|
}\quad for\quad\varphi_1^-<\varphi<\varphi _0^+,
\end{equation}

(2) when $g=0$ʱ, Eq.(\ref{eq1.3}) has the following peaked soliton
solutions
\begin{equation}
\label{eq3.3} \varphi =(c-\gamma) e^{ - |\xi| },
\end{equation}

(3) when $g_2(c)<g<0$,

(i) if $ c>\gamma$, then Eq.(\ref{eq1.3}) has the following periodic
cusp wave solutions
\begin{equation}
\label{eq3.4} u(x,t)=\varphi_3(x-ct-2nT)\quad for
\quad(2n-1)T<x-ct<(2n+1)T,
\end{equation}

(ii) if $ c<\gamma$, then Eq.(\ref{eq1.3}) has the following
periodic cusp wave solutions
\begin{equation}
\label{eq3.5} u(x,t)=\varphi_4(x-ct-2nT)\quad for\quad
(2n-1)T<x-ct<(2n+1)T,
\end{equation}
where
\begin{equation}
 \label{eq3.6}
 \beta_1(\varphi)= \frac{(2\sqrt {\varphi^2 + l_1 \varphi + l_2 }
+ 2\varphi  + l_1 )(\varphi - \varphi _0^ - )^{\alpha _1 }}{(2\sqrt
{a_1 } \sqrt {\varphi ^2 + l_1 \varphi  + l_2 } + b_1 \varphi  + l_3
)^{\alpha _1 }},
\end{equation}
\begin{equation}
\label{eq3.7} \beta_2(\varphi)= \frac{(2\sqrt {\varphi^2 + m_1
\varphi + m_2 } + 2\varphi + m_1 )(\varphi - \varphi _0^ + )^{\alpha
_2 }}{(2\sqrt {a_2 } \sqrt {\varphi ^2 + m_1 \varphi + m_2 } + b_2
\varphi + m_3 )^{\alpha _2 }},
\end{equation}
\begin{equation}
\label{eq3.8} l_1 = -\frac{3(c-\gamma)+\sqrt {(c - \gamma)^2 - 8g}
}{2},
\end{equation}
\begin{equation}
\label{eq3.9} l_2 = \frac{3(c-\gamma)^2-4g+5(c-\gamma)\sqrt {(c -
\gamma)^2 - 8g} }{8},
\end{equation}
\begin{equation}
\label{eq3.10} l_3 = \frac{(c-\gamma)^2-4g+3(c-\gamma)\sqrt {(c -
\gamma)^2 - 8g}}{2},
\end{equation}
\begin{equation}
\label{eq3.11} m_1 = -\frac{3(c-\gamma)-\sqrt {(c - \gamma)^2 - 8g}
)}{2},
\end{equation}
\begin{equation}
\label{eq3.12} m_2 = \frac{3(c-\gamma)^2-4g-5(c-\gamma)\sqrt {(c -
\gamma)^2 - 8g} }{8},
\end{equation}
\begin{equation}
\label{eq3.13} m_3 = \frac{(c-\gamma)^2-4g-3(c-\gamma)\sqrt {(c -
\gamma)^2 - 8g}}{2},
\end{equation}
\begin{equation}
\label{eq3.14} a_1 = \frac{((c-\gamma)^2-8g+3(c-\gamma)\sqrt {(c -
\gamma)^2 - 8g}}{4},
\end{equation}
\begin{equation}
\label{eq3.15} a_2 = \frac{((c-\gamma)^2-8g-3(c-\gamma)\sqrt {(c -
\gamma)^2 - 8g}}{4},
\end{equation}
\begin{equation}
\label{eq3.16} b_1 =-(c-\gamma)-\sqrt {(c - \gamma)^2 - 8g},
\end{equation}
\begin{equation}
\label{eq3.17} b_2 = -(c-\gamma)+\sqrt {(c - \gamma)^2 - 8g},
\end{equation}
\begin{equation}
\label{eq3.18} \alpha _1 = - \frac{3(c-\gamma)+\sqrt {(c - \gamma)^2
- 8g} }{2\sqrt {(c - \gamma)^2 - 8g + 3(c-\gamma)\sqrt {(c -
\gamma)^2 - 8g} } },
\end{equation}
\begin{equation}
\label{eq3.19} \alpha _2 = \frac{-3(c-\gamma)+\sqrt {(c - \gamma)^2
- 8g} }{2\sqrt {(c - \gamma)^2 - 8g -3(c-\gamma)\sqrt {(c -
\gamma)^2 - 8g} } },
\end{equation}
\begin{equation}
\label{eq3.20}
 \varphi_3
 (\xi)=l_+
 e^{-|\xi|}+l_-e^{|\xi|}
\quad for \quad \sqrt{-g}\leq \varphi_3 \leq c-\gamma,
\end{equation}
\begin{equation}
\label{eq3.21}
 \varphi_4
 (\xi)=l_+
 e^{|\xi|}+l_-e^{-|\xi|}
\quad for \quad c-\gamma\leq \varphi_4 \leq -\sqrt{-g} ,
\end{equation}
\begin{equation}
\label{eq3.22} l_\pm=\frac{c-\gamma\pm \sqrt{(c-\gamma)^2+g}}{2},
\end{equation}
\begin{equation}
\label{eq3.23} T=\left |\ln(\sqrt{-g}+\sqrt{-2g}) -\ln(2l_{-})\right
|.
\end{equation}
\begin{equation}
\label{eq3.24} \varphi_1^\pm=\frac{3}{4}(c-\gamma)\pm \frac{1}{4}
 \sqrt {(c - \gamma)^2 - 8g} \mp \frac{1}{2}\sqrt {(c - \gamma)^2 \mp (c-\gamma)\sqrt {(c -
\gamma)^2 - 8g} },
\end{equation}
 $\varphi _0^+$ and $\varphi _0^-$ are as in (\ref{eq2.11}).

\end{theorem}

\begin{figure}[h]
\centering \subfloat[]{\label{fig:3.6}
\includegraphics[height=1.2in,width=2.3in]{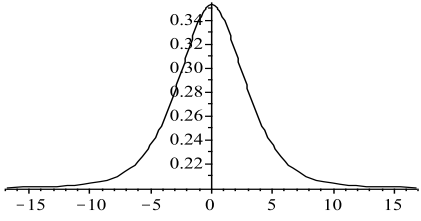}}\hspace{0.1\textwidth}
\subfloat[]{ \label{fig:b}
\includegraphics[height=1.2in,width=2.3in]{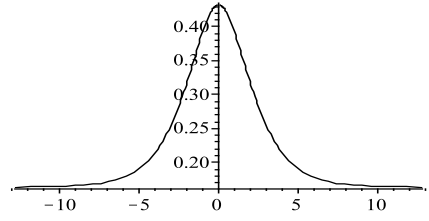}}\\
\subfloat[]{ \label{fig:c}
\includegraphics[height=1.2in,width=2.3in]{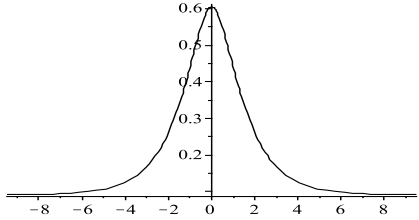}}\hspace{0.1\textwidth}
\subfloat[]{ \label{fig:d}
\includegraphics[height=1.2in,width=2.3in]{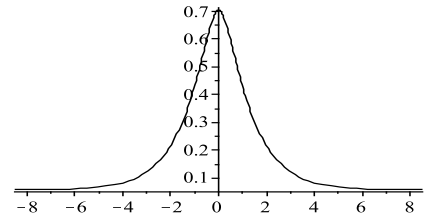}}\\
 \caption{Smooth hump-like soliton solutions of Eq.(\ref{eq1.3})( $c = 2$,
$\gamma=1$). (a) $g = 0.12$; (b)g=0.1; (c) g=0.075; (d)
g=0.05.}\label{f3}
\end{figure}

\begin{figure}[h]
\centering \subfloat[]{\label{fig:a}
\includegraphics[height=1.1in,width=2.3in]{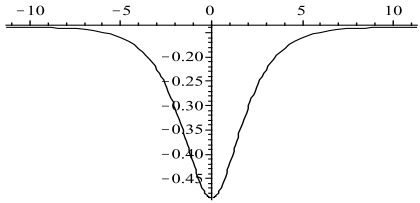}}\hspace{0.1\textwidth}
\subfloat[]{ \label{fig:b}
\includegraphics[height=1.1in,width=2.3in]{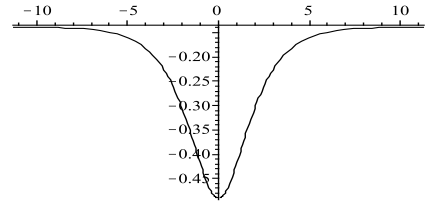}}\\
\subfloat[]{ \label{fig:c}
\includegraphics[height=1.1in,width=2.3in]{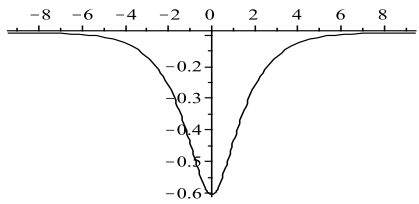}}\hspace{0.1\textwidth}
\subfloat[]{ \label{fig:d}
\includegraphics[height=1.1in,width=2.3in]{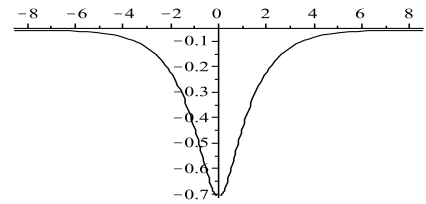}}
 \caption{Smooth valley-like soliton  solutions of Eq.(\ref{eq1.3}).($c = 1$,
$\gamma=2$). (a) g = 0.12; (b) g=0.1; (c) g=0.075; (d)
g=0.05.}\label{f4}
\end{figure}

\begin{figure}[h]
\centering \subfloat[]{\label{fig:a}
\includegraphics[height=1.2in,width=2.1in]{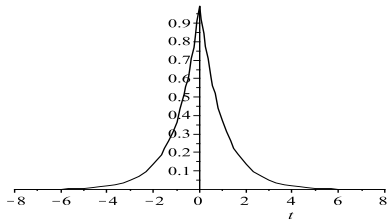}}\hspace{0.1\textwidth}
\subfloat[]{ \label{fig:b}
\includegraphics[height=1.2in,width=2.1in]{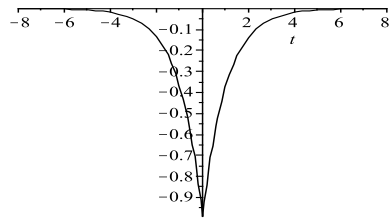}}
 \caption{Peaked soliton solutions of Eq.(\ref{eq1.3}). (a) $c = 2$,
$\gamma=1$; (b) $c=1$, $\gamma=2$.}\label{f5}
\end{figure}

\begin{figure}[h]
\centering \subfloat[]{\label{fig:a}
\includegraphics[height=1.1in,width=2.in]{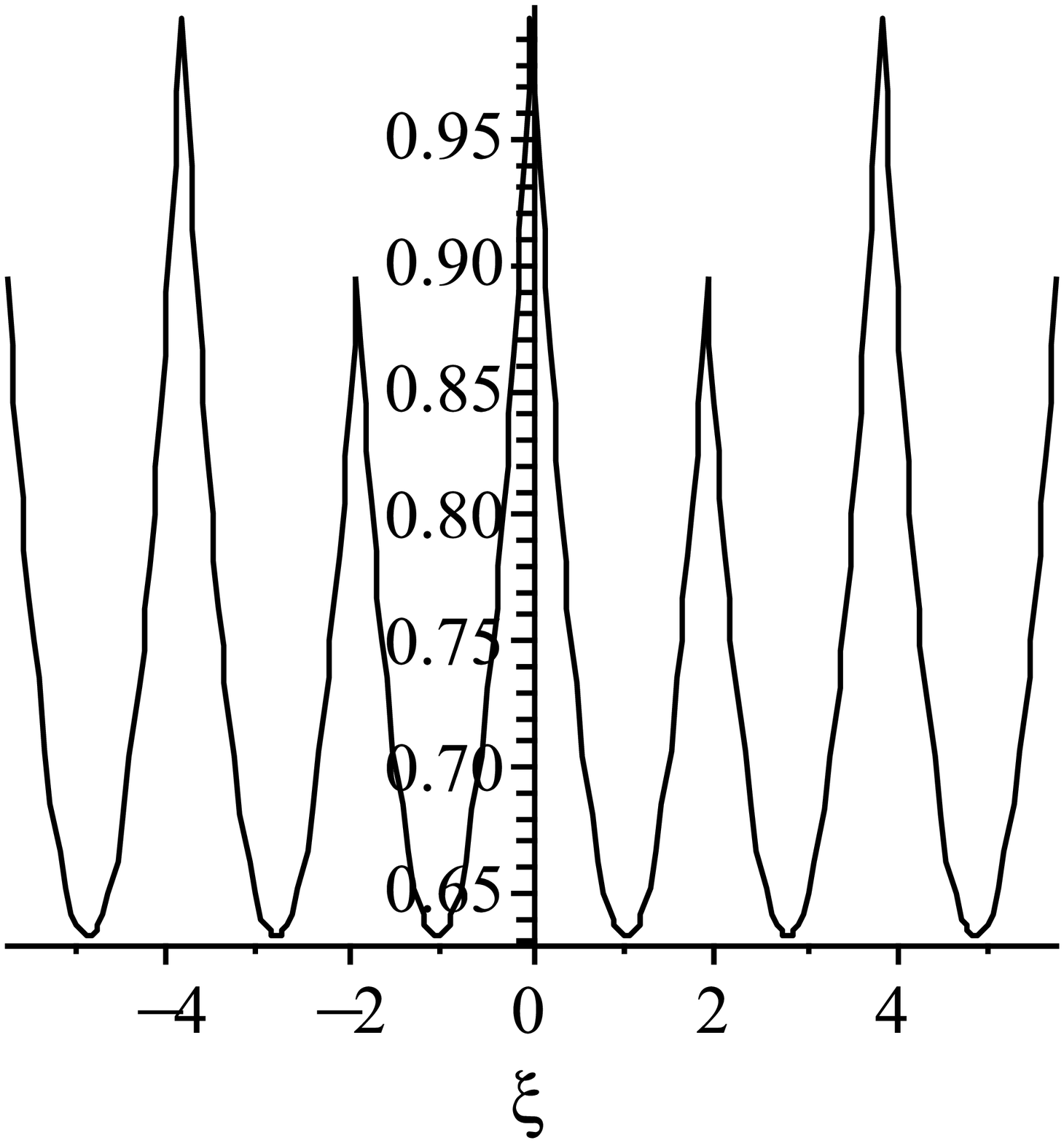}}\hspace{0.15\textwidth}
\subfloat[]{ \label{fig:b}
\includegraphics[height=1.1in,width=2.in]{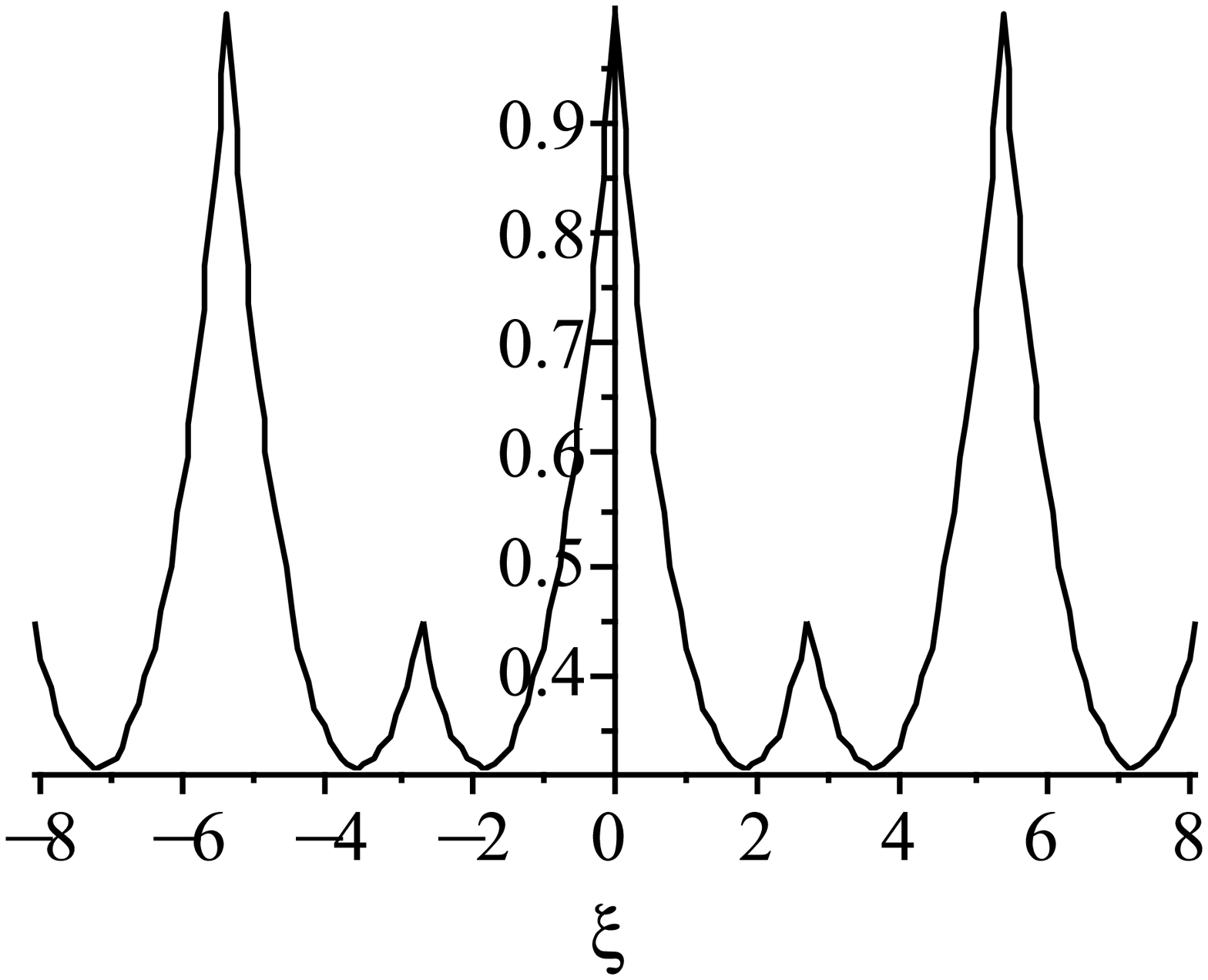}}\\
\subfloat[]{ \label{fig:c}
\includegraphics[height=1.2in,width=2.1in]{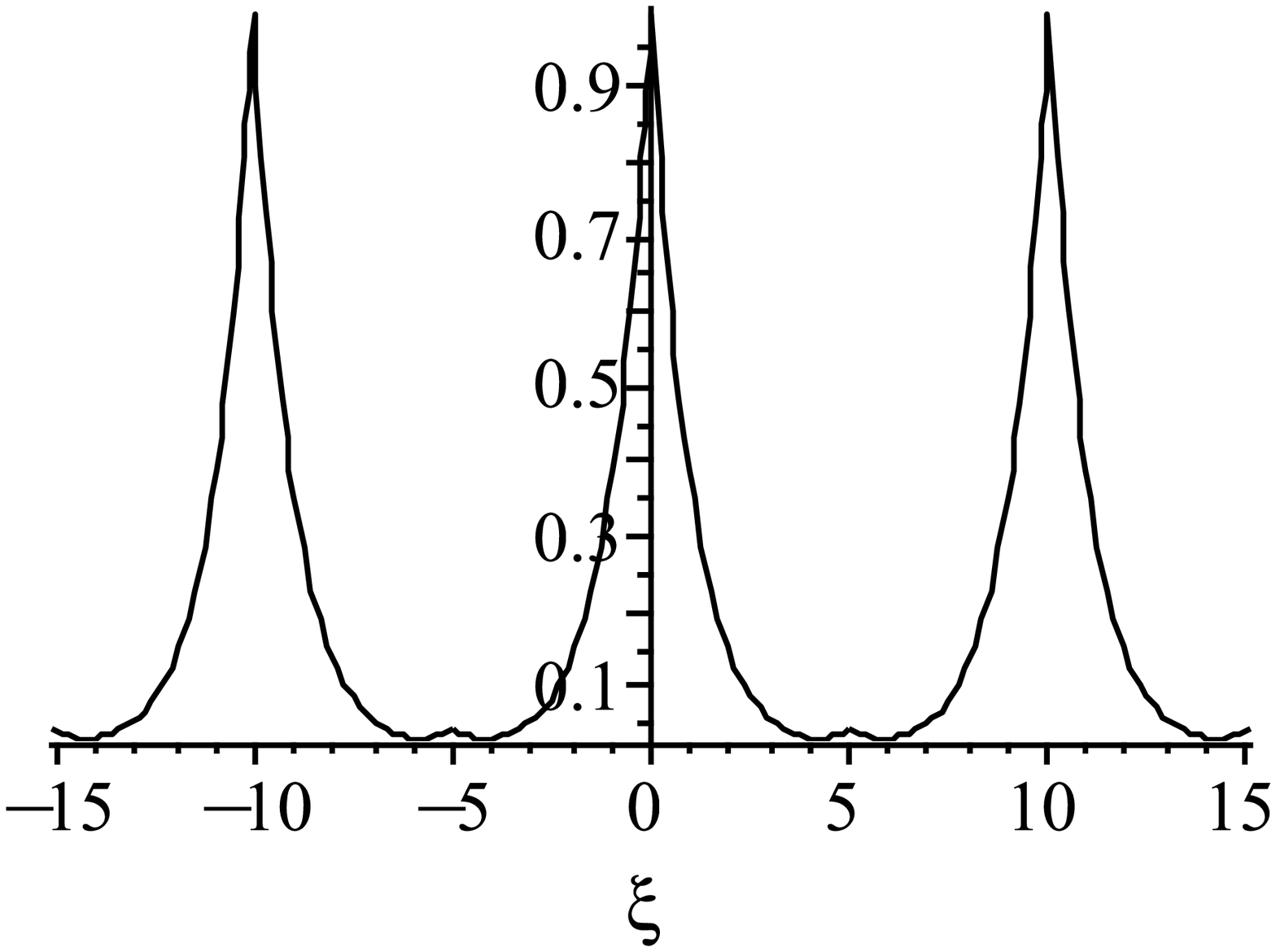}}\hspace{0.15\textwidth}
\subfloat[]{ \label{fig:d}
\includegraphics[height=1.2in,width=2.1in]{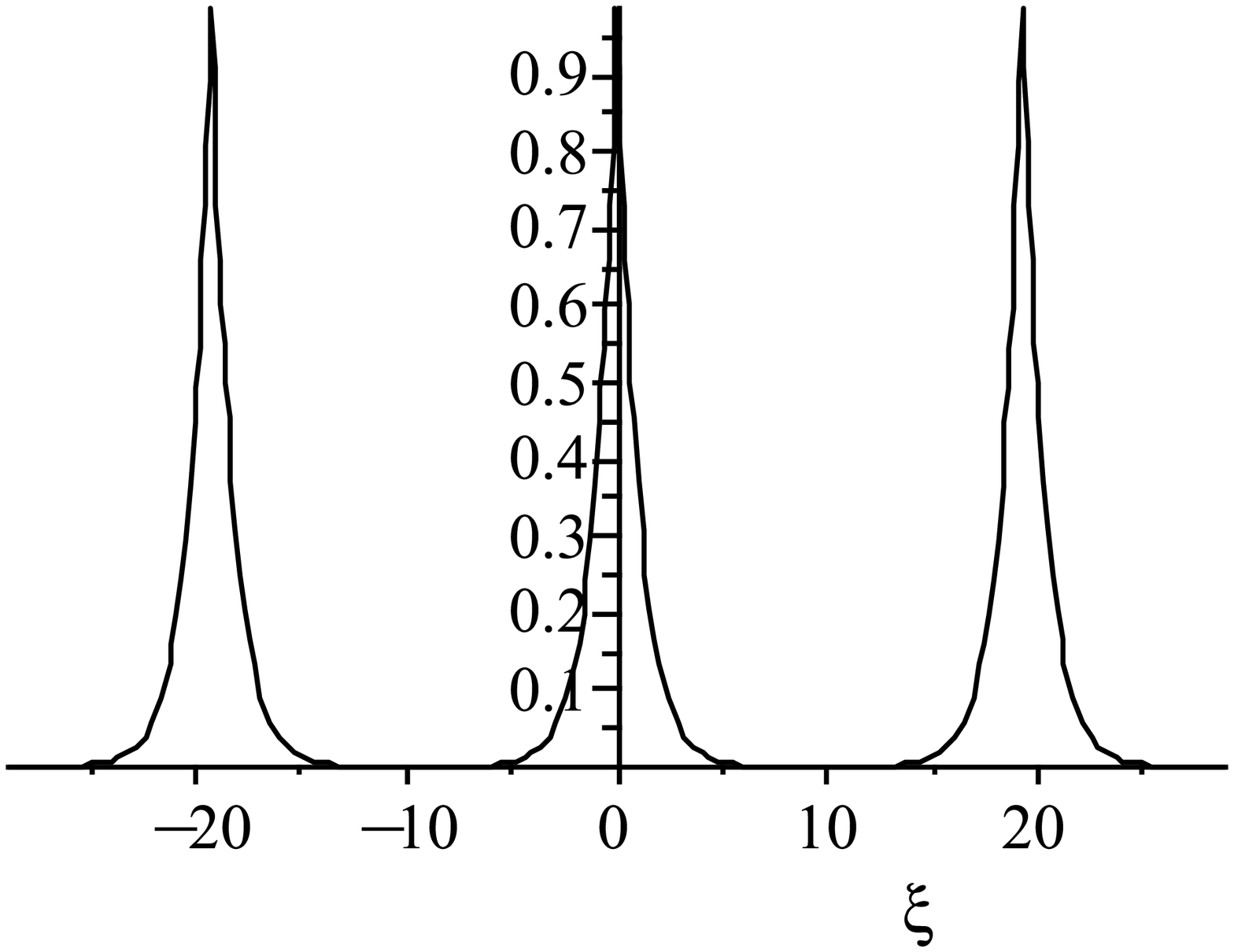}}\\
 \caption{Period cuspon solutions of Eq.(\ref{eq1.3})( $c = 2$,
$\gamma=1$). (a) g = -0.4; (b) g=-0.1; (c) g=-0.01; (d)
g=-0.0000001.}\label{f6}
\end{figure}

\begin{figure}[h]
\centering \subfloat[]{\label{fig:a}
\includegraphics[height=1.1in,width=2.in]{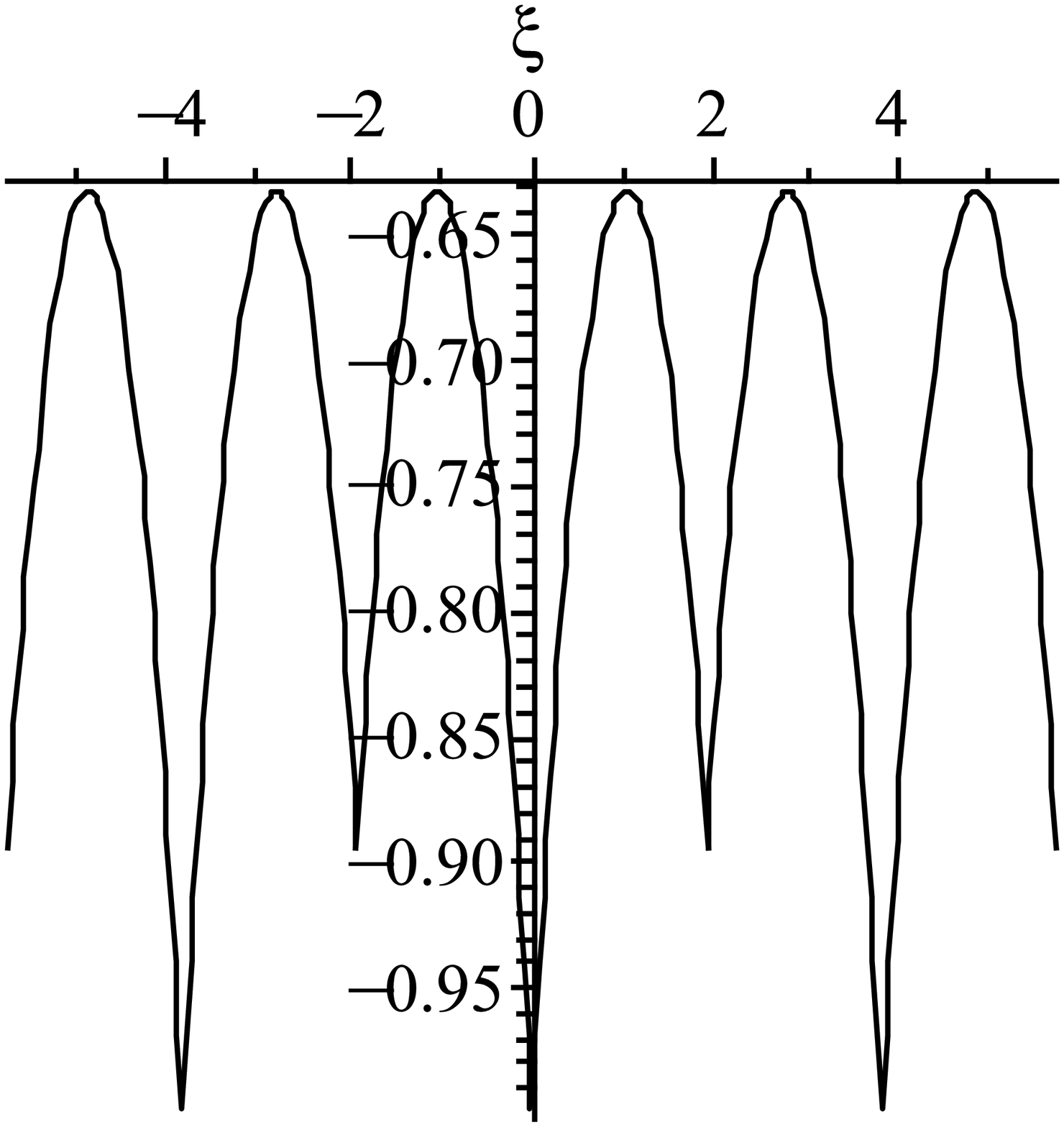}}\hspace{0.15\textwidth}
\subfloat[]{ \label{fig:b}
\includegraphics[height=1.1in,width=2.in]{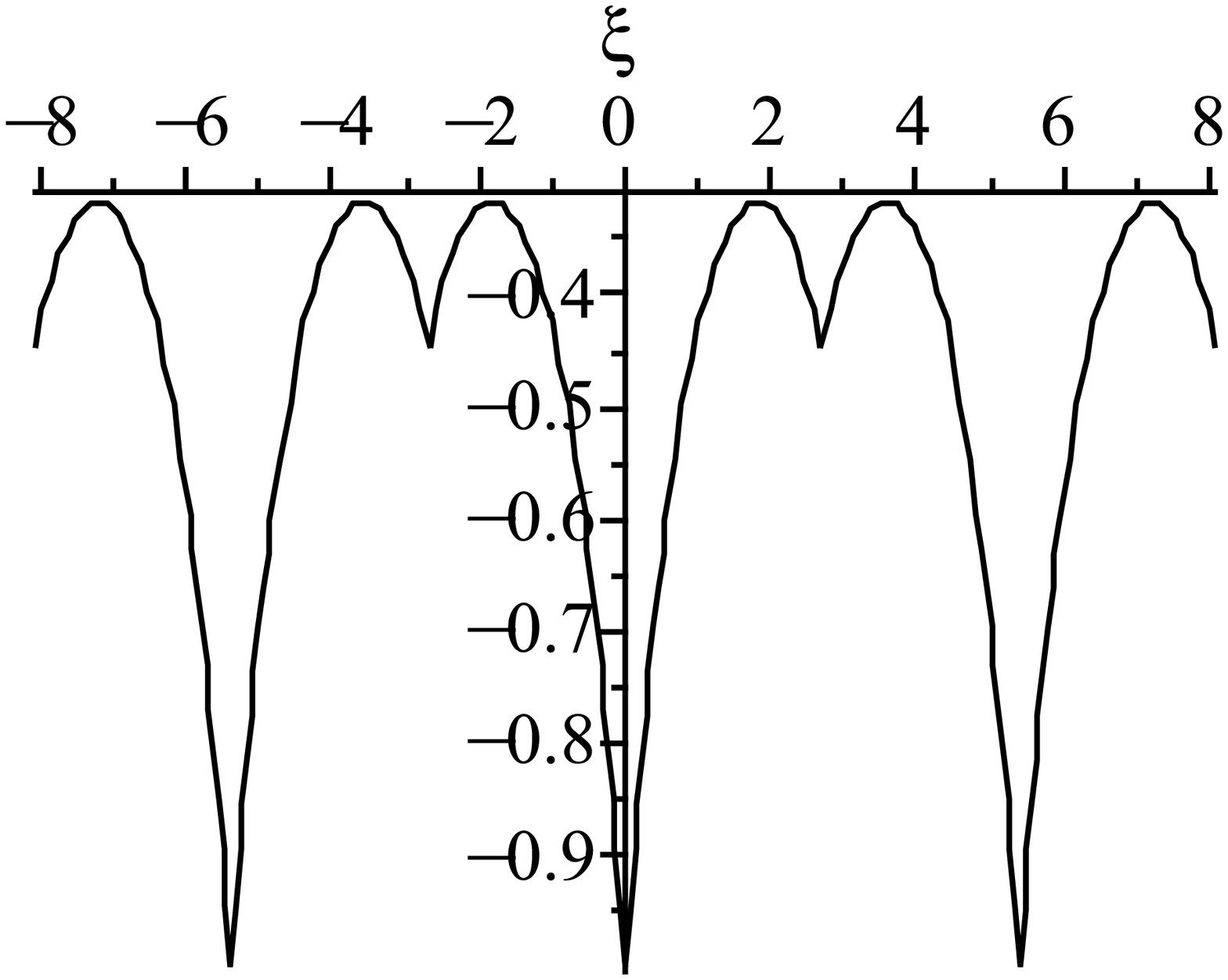}}\\
\subfloat[]{ \label{fig:c}
\includegraphics[height=1.2in,width=2.in]{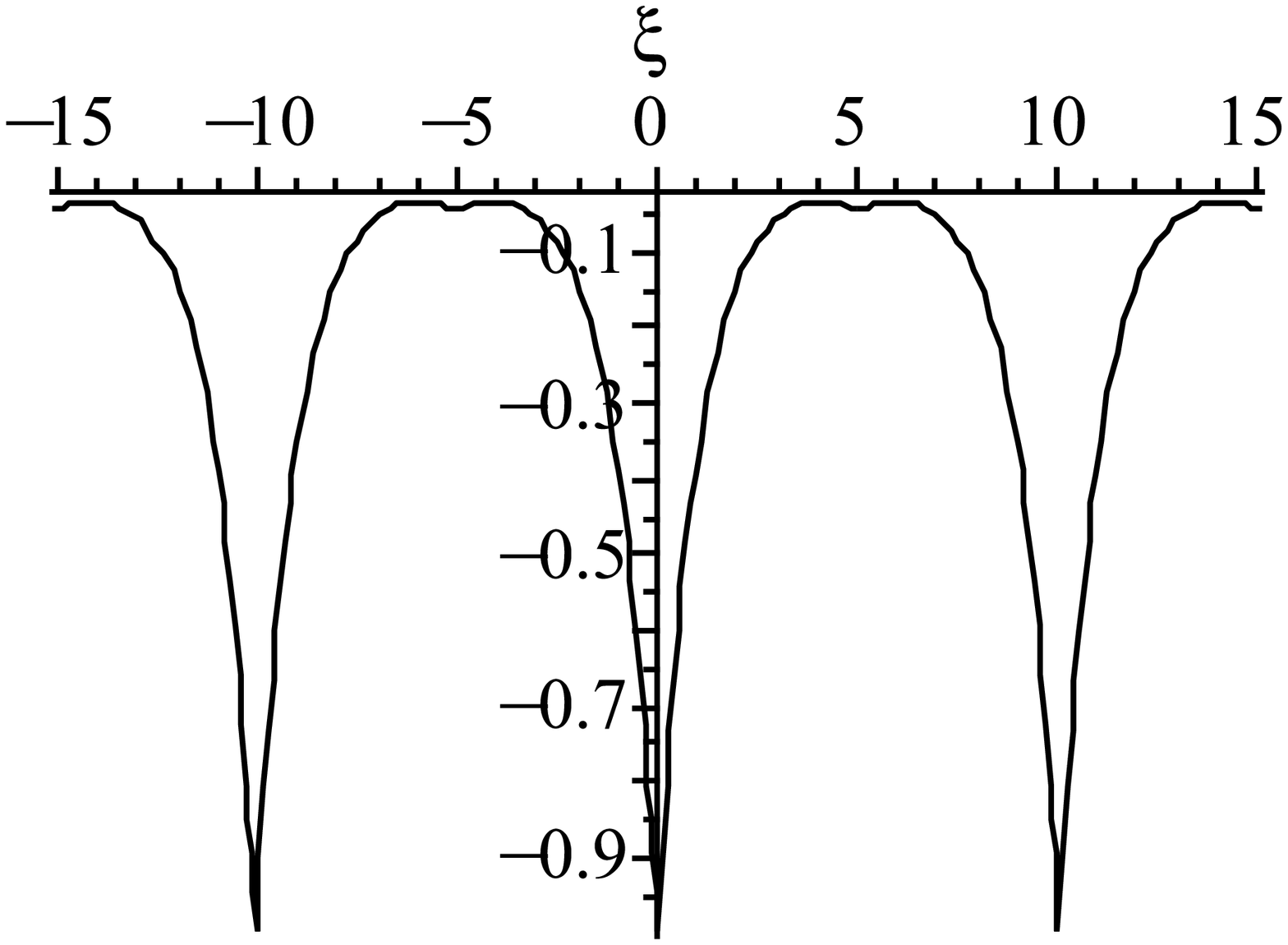}}\hspace{0.15\textwidth}
\subfloat[]{ \label{fig:d}
\includegraphics[height=1.in,width=2.in]{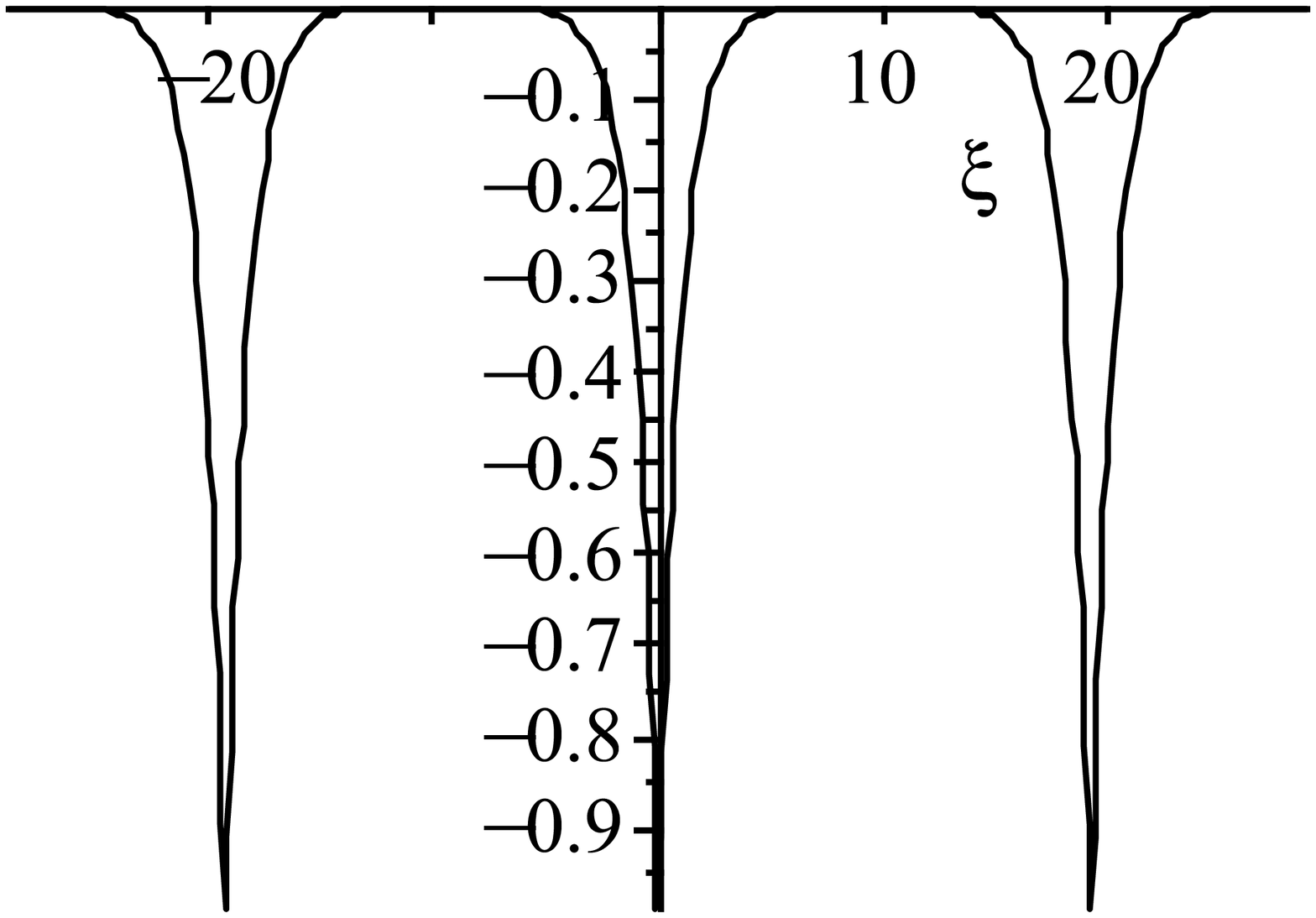}}\\
 \caption{Period cuspon solutions of Eq.(\ref{eq1.3})($c = 1$,
$\gamma=2$). (a) g = -0.4; (b) g=-0.1; (c) g=-0.01; (d)
g=-0.0000001.}\label{f7}
\end{figure}

Before proving this theorem, we take a set of data and employ Maple
to display the graphs of smooth solion, peaked soliton and periodic
cuspon solutions of Eq.(\ref{eq1.3}), see Fig.\ref{f3}-Fig.\ref{f7}.

\begin{figure}[h]
\centering
\subfloat[]{\includegraphics[height=1.2in,width=1.4in]{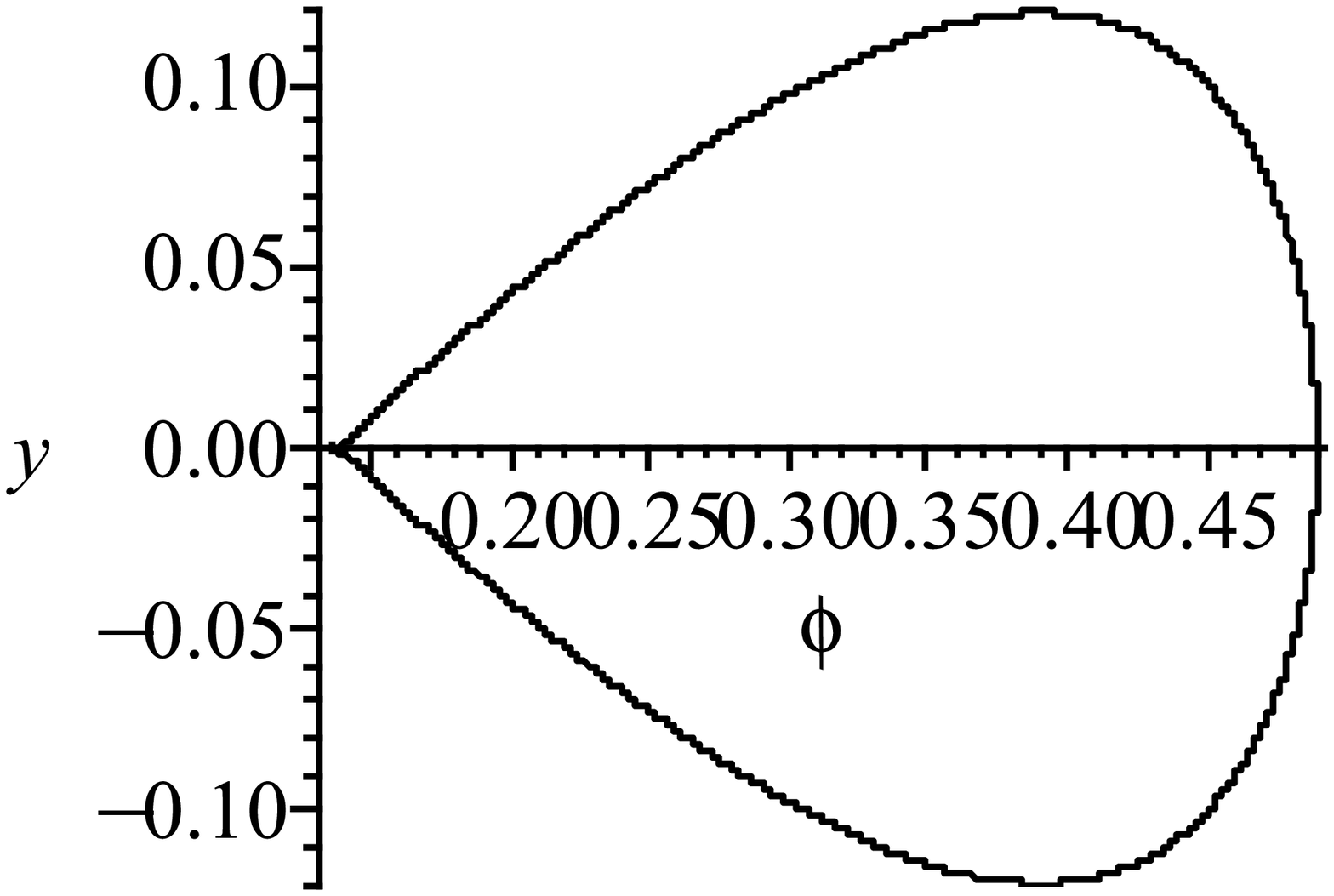}}\hspace{0.08\textwidth}
\subfloat[]{\includegraphics[height=1.2in,width=1.4in]{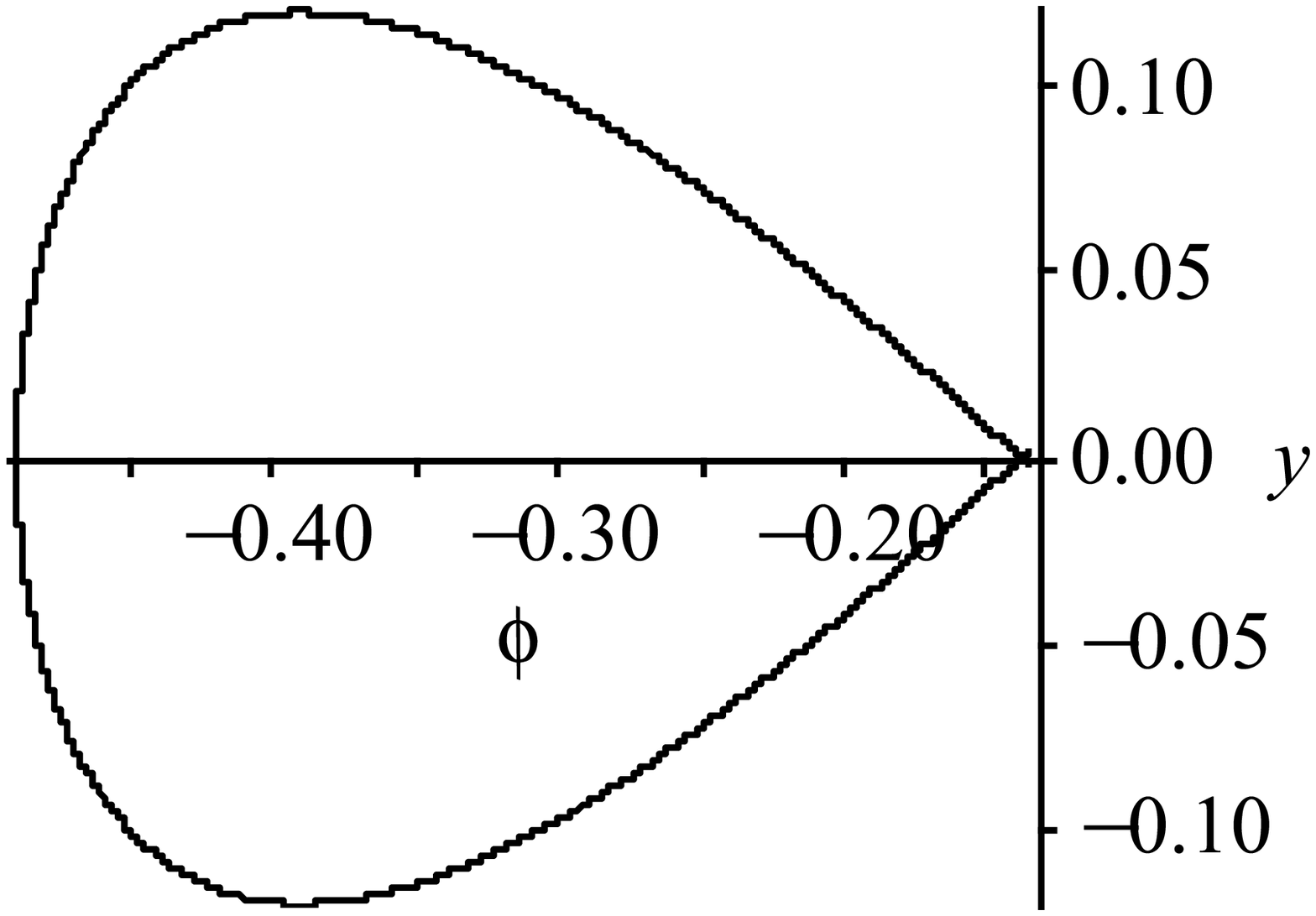}}\hspace{0.08\textwidth}
\subfloat[]{\includegraphics[height=1.2in,width=1.4in]{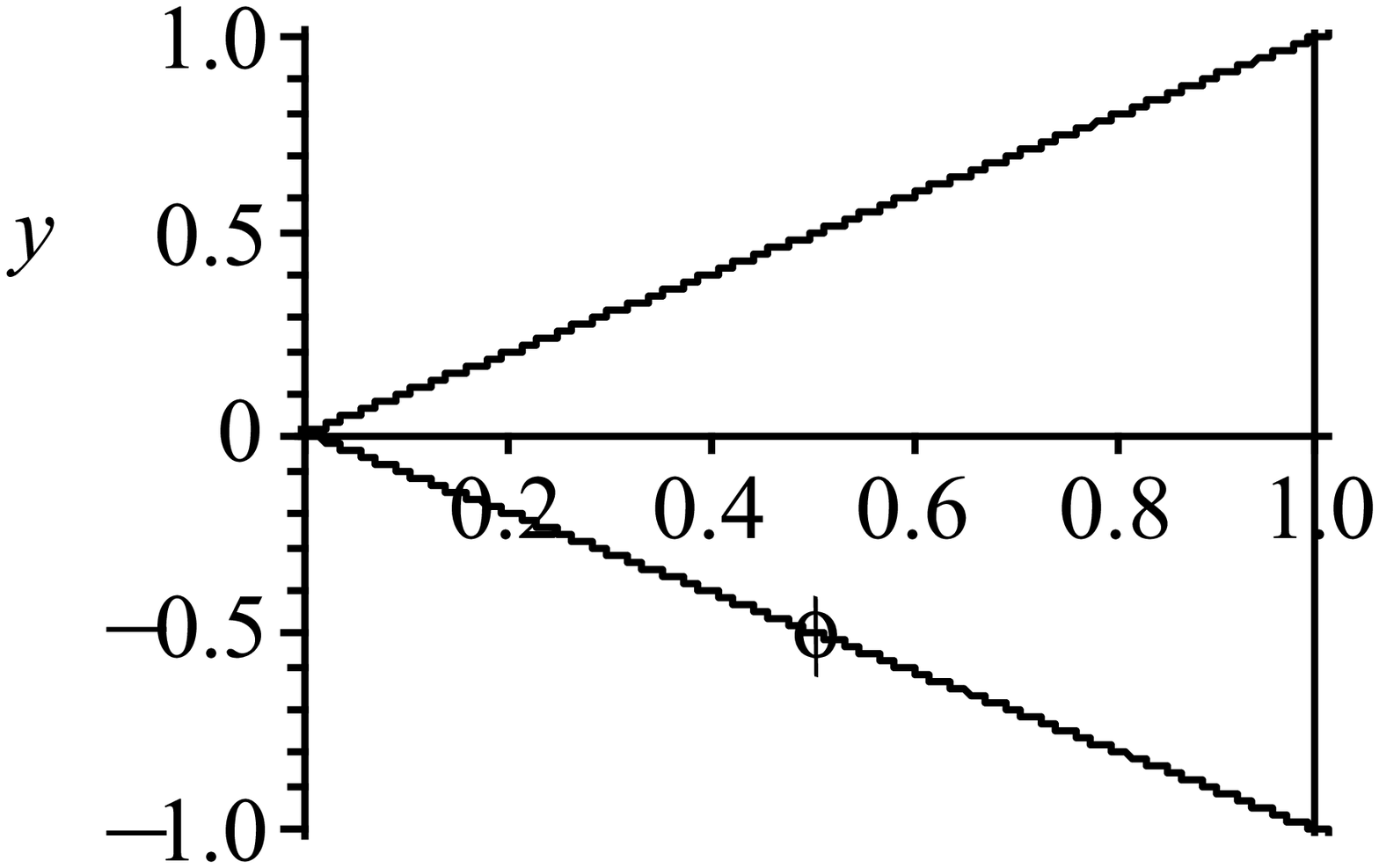}}\\\hspace{0.02\textwidth}
\subfloat[]{\includegraphics[height=1.2in,width=1.2in]{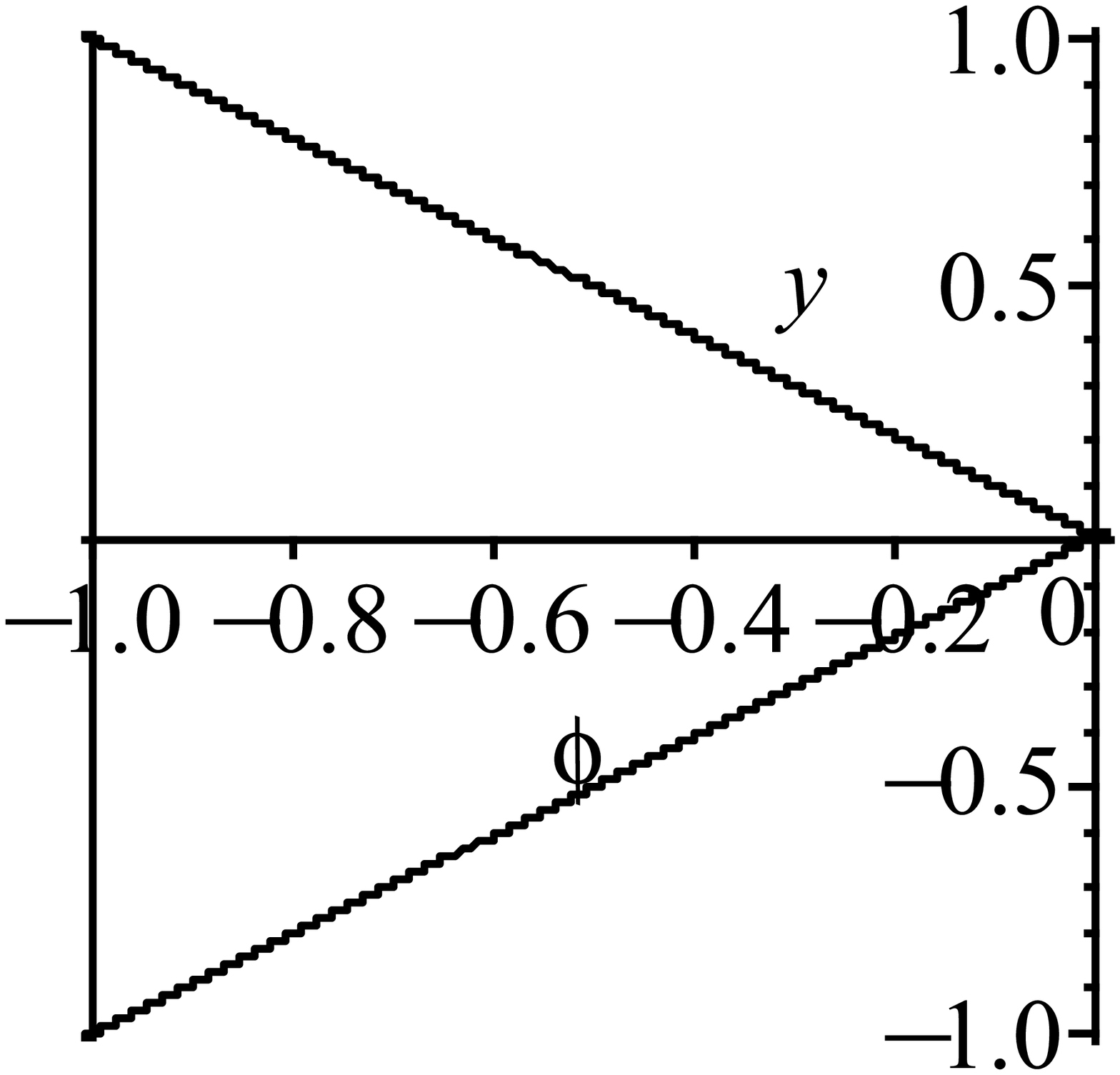}}\hspace{0.09\textwidth}
\subfloat[]{\includegraphics[height=1.2in,width=1.4in]{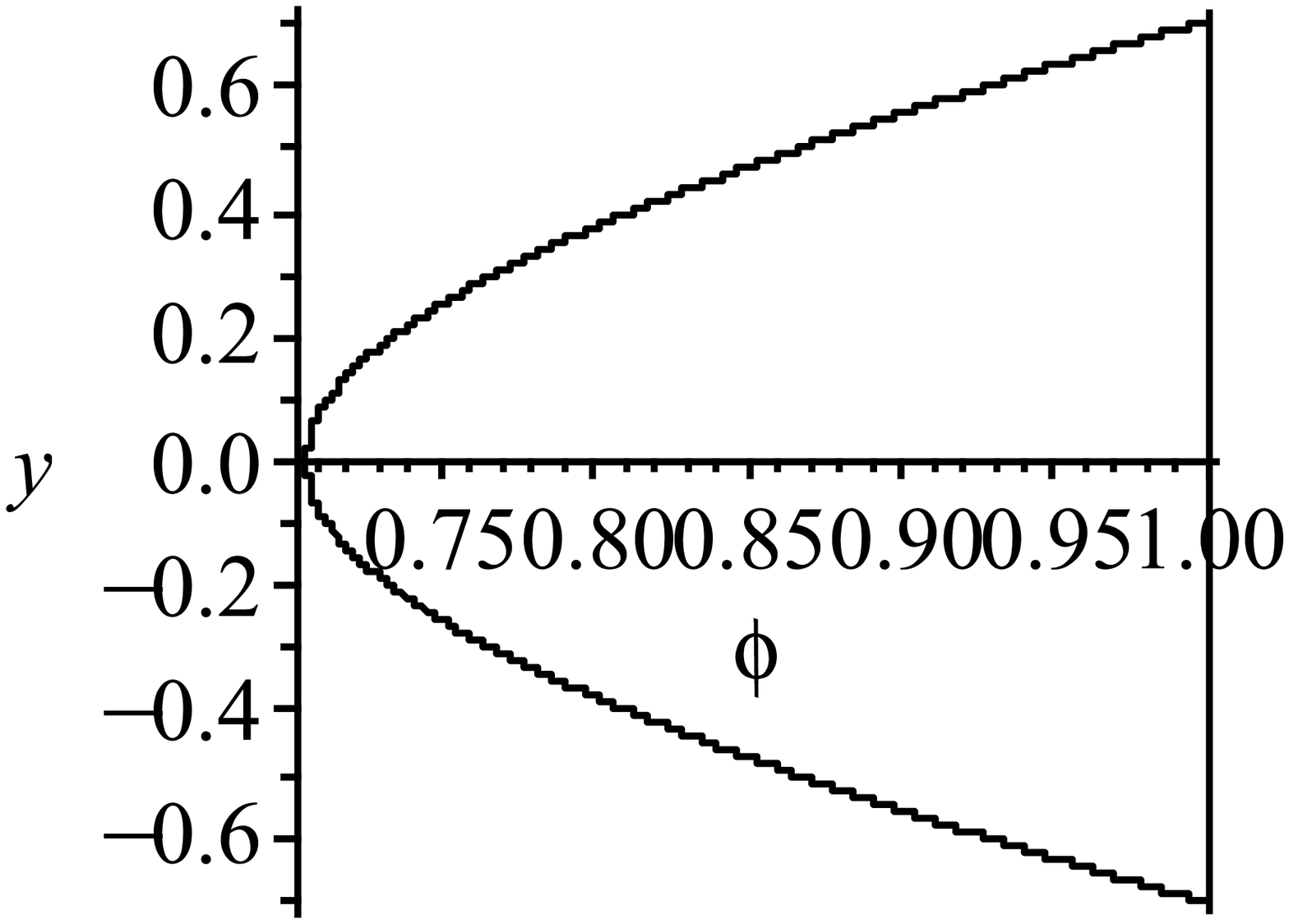}}\hspace{0.1\textwidth}
\subfloat[]{\includegraphics[height=1.2in,width=1.4in]{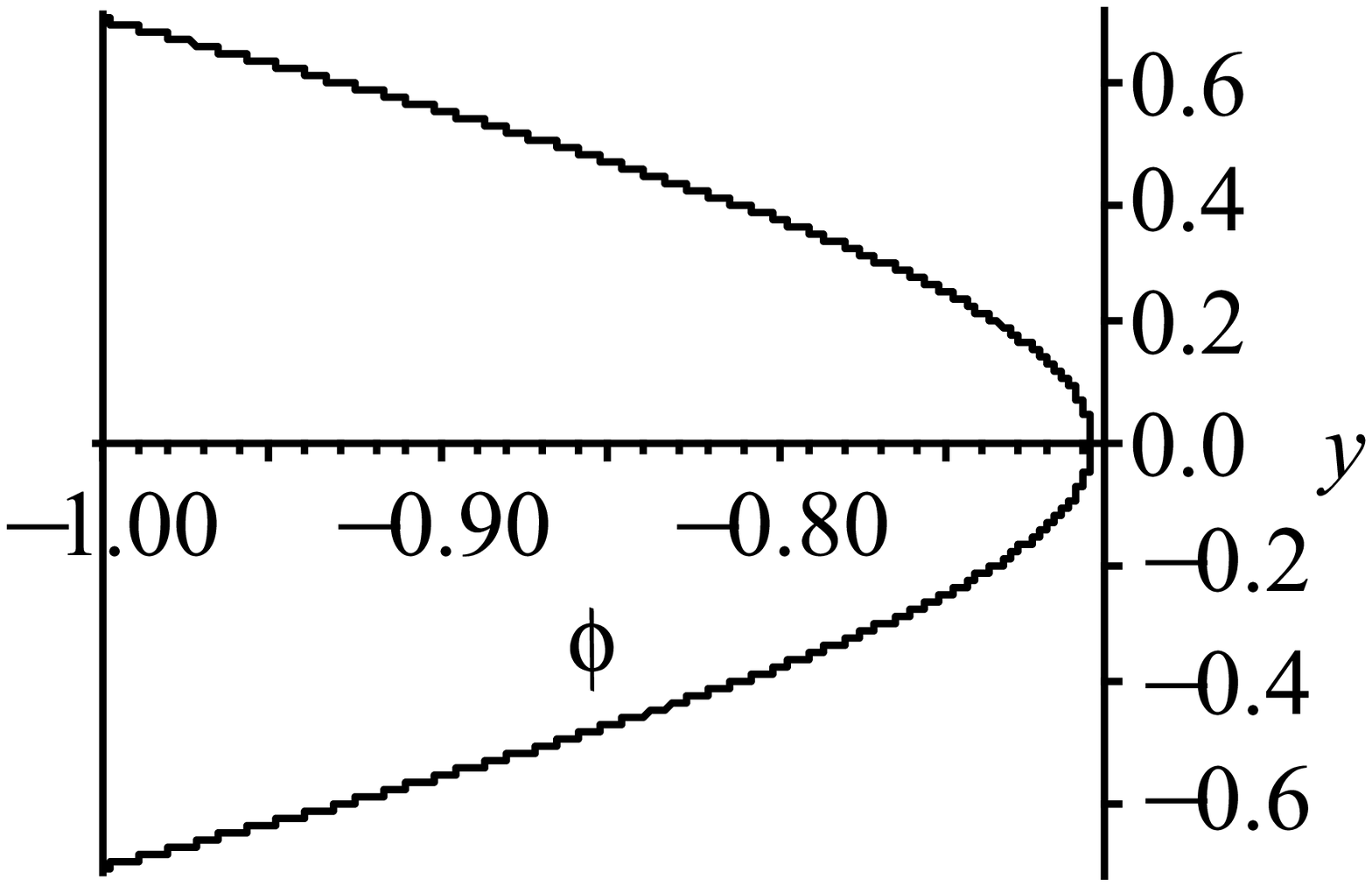}}
 \caption{The orbits of system (\ref{eq2.5}) connecting with the saddle points. (a)$ 0< g<g_1 (c) ,
c>\gamma$; (b) $0< g<g_1 (c), c<\gamma$; (c) $g =0, c>\gamma$;  (d)
$g=0, c<\gamma$; (e) $g_2(c)<g<0, c>\gamma$; (f) $g_2(c)<g<0,
c<\gamma$.}\label{f8}
\end{figure}

\begin{proof}
Usually, a solion solution of Eq.(\ref{eq1.3}) corresponds to a
homoclinic orbit of system (\ref{eq2.5}) and a periodic travelling
wave solution of Eq.(\ref{eq1.3}) corresponds to a periodic orbit of
system (\ref{eq2.5}).  The graphs of homoclinic orbit, periodic
orbit of system (\ref{eq2.5}) and their limit curves are shown in
Fig.\ref{f8}.

(1) When $ 0< g<g_1(c), c>\gamma$, system (\ref{eq2.5}) has a
homoclinic orbit (see Fig.\ref{f8}(a)). This homoclinic orbit can be
expressed as
\begin{equation}
\label{eq3.25}  y =\pm \frac{(\varphi - \varphi _0^ - )\sqrt
{\varphi ^2 + l_1 \varphi + l_2 } }{c -\gamma- \varphi }\quad
for\quad \varphi _0^ -<\varphi<\varphi_1^+.
\end{equation}
Substituting  (\ref{eq3.25}) into the first equation of system
(\ref{eq2.3}) and integrating along this homoclinic orbit, we obtain
(\ref{eq3.1}).

When $ 0< g<g_1 (c), c<\gamma$, we can obtain (\ref{eq3.2}) in
similar way.

(2) When $g=0, c>\gamma$, system (\ref{eq2.5}) has a homoclinic
orbit that consists of the following three line segments (see
Fig.\ref{f8}(c)).
\begin{equation}
\label{eq3.26} y=\pm \varphi \quad for \quad \varphi_0^-\leq
\varphi\leq \varphi_1^+ ,
\end{equation}
and
\begin{equation}
\label{eq3.27} \varphi=c-\gamma \quad for \quad
-\sqrt{(c-\gamma)^2+g}\leq y\leq \sqrt{(c-\gamma)^2+g}.
\end{equation}
Substituting  (\ref{eq3.26}) into the first equation of system
(\ref{eq2.3}) and integrating along this orbit, we obtain
(\ref{eq3.3}).

When $g=0, c<\gamma$, we can also obtain (\ref{eq3.3}).

(3) When $g_2(c)<g<0, c>\gamma$, system (\ref{eq2.5}) has a periodic
orbit (see Fig. \ref{f8}(e)). This periodic orbit can be expressed
as
\begin{equation}
\label{eq3.28} y=\pm\sqrt{\varphi^2+g} \quad for \quad
\sqrt{-g}\leq \varphi \leq c-\gamma,
\end{equation}
and
\begin{equation}
\label{eq3.29} \varphi=c-\gamma \quad for \quad
-\sqrt{(c-\gamma)^2+g}\leq y\leq \sqrt{(c-\gamma)^2+g}.
\end{equation}
Substituting  (\ref{eq3.28}) into the first equation of system
(\ref{eq2.3}) and integrating along this periodic orbit, we obtain
(\ref{eq3.4}).

When $g_2(c)<g<0, c<\gamma$, we can obtain (\ref{eq3.5}).
\end{proof}
\begin{remark}
From the above discussion, we can see that when $g<0, g\rightarrow
0$, the period of the periodic cusp wave solution becomes bigger and
bigger, and the periodic cuspon solutions (\ref{eq3.4}) and
(\ref{eq3.5}) tend to the peaked soliton solutions (\ref{eq3.3}) .
When $g>0, g\rightarrow 0$, the smooth hump-like soliton solutions
(\ref{eq3.1}) and  the smooth valley-like soliton solutions
(\ref{eq3.2}) lose their smoothness and tend to the peaked soliton
solutions (\ref{eq3.3}).
\end{remark}

\section{Discussion}
In this paper, we obtain the solitons, peakons, and periodic cuspons
of a generalized Degasperis-Procesi equation (\ref{eq1.3}). These
solitons denote the nonlinear localized waves on the shallow water's
free surface that retain their individuality under interaction and
eventually travel with their original shapes and speeds. The balance
between the nonlinear steepening and dispersion effect under
Eq.(\ref{eq1.3}) gives rise to these solitons.

The peakon travels with speed equal to its peak amplitude. This
solution is nonanalytic, having a jump in derivative at its peak.
Peakons are true solitons that interact via elastic collisions under
Eq.(\ref{eq1.3}). We claim that the existence of a singular straight
line for the planar dynamical system (\ref{eq2.3}) is the original
reason why the travelling waves lose their smoothness.

Also, the periodic cuspon solution is nonanalytic, having a jump in
derivative at its each cusp.

\end{document}